\documentclass[aps,prl,reprint,superscriptaddress,floatfix]{revtex4-2}
\usepackage{amsmath}
\usepackage{graphicx}
\usepackage{lmodern}
\usepackage{amsmath}
\usepackage{color} 
\usepackage{amssymb}
\usepackage{bm}
\usepackage{braket}
\usepackage{mathtools}

\newcommand{\bk}{\bm{k}}
\newcommand{\bp}{\bm{p}}
\newcommand{\bq}{\bm{q}}

\DeclareMathOperator{\Imm}{Im}

\newcommand{\YMW}[1]{\textcolor{black}{#1}}

\usepackage[dvipsnames]{xcolor}
\usepackage[colorlinks=true]{hyperref}
\hypersetup{
    colorlinks=true,
    linkcolor=blue,
    citecolor=blue,
    urlcolor=blue
} 

\usepackage[compat=1.1.0]{tikz-feynman}

\makeatletter
\def\maketitle{
\@author@finish
\title@column\titleblock@produce
\suppressfloats[t]}
\makeatother

\usepackage{tikz}
\usetikzlibrary{decorations.pathreplacing,decorations.markings}
\tikzset{
  on each segment/.style={
    decorate,
    decoration={
      show path construction,
      moveto code={},
      lineto code={
        \path [#1]
        (\tikzinputsegmentfirst) -- (\tikzinputsegmentlast);
      },
      curveto code={
        \path [#1] (\tikzinputsegmentfirst)
        .. controls
        (\tikzinputsegmentsupporta) and (\tikzinputsegmentsupportb)
        ..
        (\tikzinputsegmentlast);
      },
      closepath code={
        \path [#1]
        (\tikzinputsegmentfirst) -- (\tikzinputsegmentlast);
      },
    },
  },
  mid arrow/.style={postaction={decorate,decoration={
        markings,
        mark=at position .5 with {\arrow[#1]{stealth}}
      }}},
}

\newcommand{\be} {\begin{equation}}
\newcommand{\ee} {\end{equation}}

\begin{document}
\title{Intra-unit-cell singlet pairing mediated by altermagnetic fluctuations}
\author{Yi-Ming Wu}
\affiliation{Stanford Institute for Theoretical Physics, Stanford University, Stanford, California 94305, USA}
\email[]{yimwu@stanford.edu}
\author{Yuxuan Wang}
\affiliation{Department of Physics, University of Florida, Gainesville, Florida 32611, USA}
\email[]{yuxuan.wang@ufl.edu}
\author{Rafael M. Fernandes}
\affiliation{Department of Physics, The Grainger College of Engineering, University of Illinois Urbana-Champaign, Urbana, Illinois 61801, USA}
\affiliation{Anthony J. Leggett Institute for Condensed Matter Theory, The Grainger College of Engineering, University of Illinois Urbana-Champaign, Urbana, Illinois 61801, USA}
\email[]{rafaelf@illinois.edu}

\begin{abstract}
    {We investigate the superconducting instabilities induced by altermagnetic fluctuations. Because of the non-trivial sublattice structure of the altermagnetic order,  shorter-range and longer-range fluctuations favor qualitatively different types of pairing states. Specifically, while the latter stabilize a standard spin-triplet $p$-wave state, just like ferromagnetic fluctuations, the former leads to intra-unit-cell pairing, in which the Cooper pairs are formed by electrons from different sublattices. The symmetry of the intra-unit-cell gap function can be not only $p$-wave, but also spin-singlet $s$-wave and $d$-wave, depending on the shape of the Fermi surface. We also show that coexistence with altermagnetic order promotes intrinsic non-trivial topology, such as protected Bogoliubov Fermi surfaces and higher-order topological superconductivity. Our work establishes the key role played by sublattice degrees of freedom in altermagnetic-fluctuation mediated interactions.}
\end{abstract}

\maketitle

\date{\today}

 {\it Introduction.}-- A key paradigm of unconventional superconductivity is pairing mediated by the exchange of soft collective electronic modes~\cite{ScalapinoRMP2012,Chubukov2002spinfluctuationmodeldwave,Metlitski1,Metlitski2,Chubukov_Schmalian,Senthil2015,Hirschfeld2011,WangChubukovPRL2013,LedererPRL2015,WangPRB2015,KoziiFuPRL2015,wang2016topological,WangShattnerPRB2017,Lederer2017,WangPRB2021,WuWangPRB2022,WangChubukov2025}. The symmetry of the resulting superconducting order depends on both the degree of freedom involved (e.g., charge, spin, valley) and the wave-vector in which the fluctuations are peaked \cite{palle2024superconductivity}. For instance, while fluctuations of uniform order in the charge channel (such as nematic~\cite{LedererPRL2015,Klein2018} or ferroelectric~\cite{Gastiasoro2020}) generally favor $s$-wave spin-singlet pairing, even-parity fluctuations in the spin channel (such as ferromagnetic~\cite{LonzarichPRB1999,Millis2001,Chubukov2003,LeeWenPRB2008,WuPRB2023}) promote $p$-wave spin triplet pairing. \YMW{When the spin order is spatially staggered with a finite wave-vector $\bm Q$, such as antiferromagnetism, fluctuations can mediate $d$-wave spin-singlet pairing~\cite{ScalapinoRMP2012,Chubukov2002spinfluctuationmodeldwave,Metlitski1,WangChubukovPRL2013}.}

\begin{figure}
    \includegraphics[width=8cm]{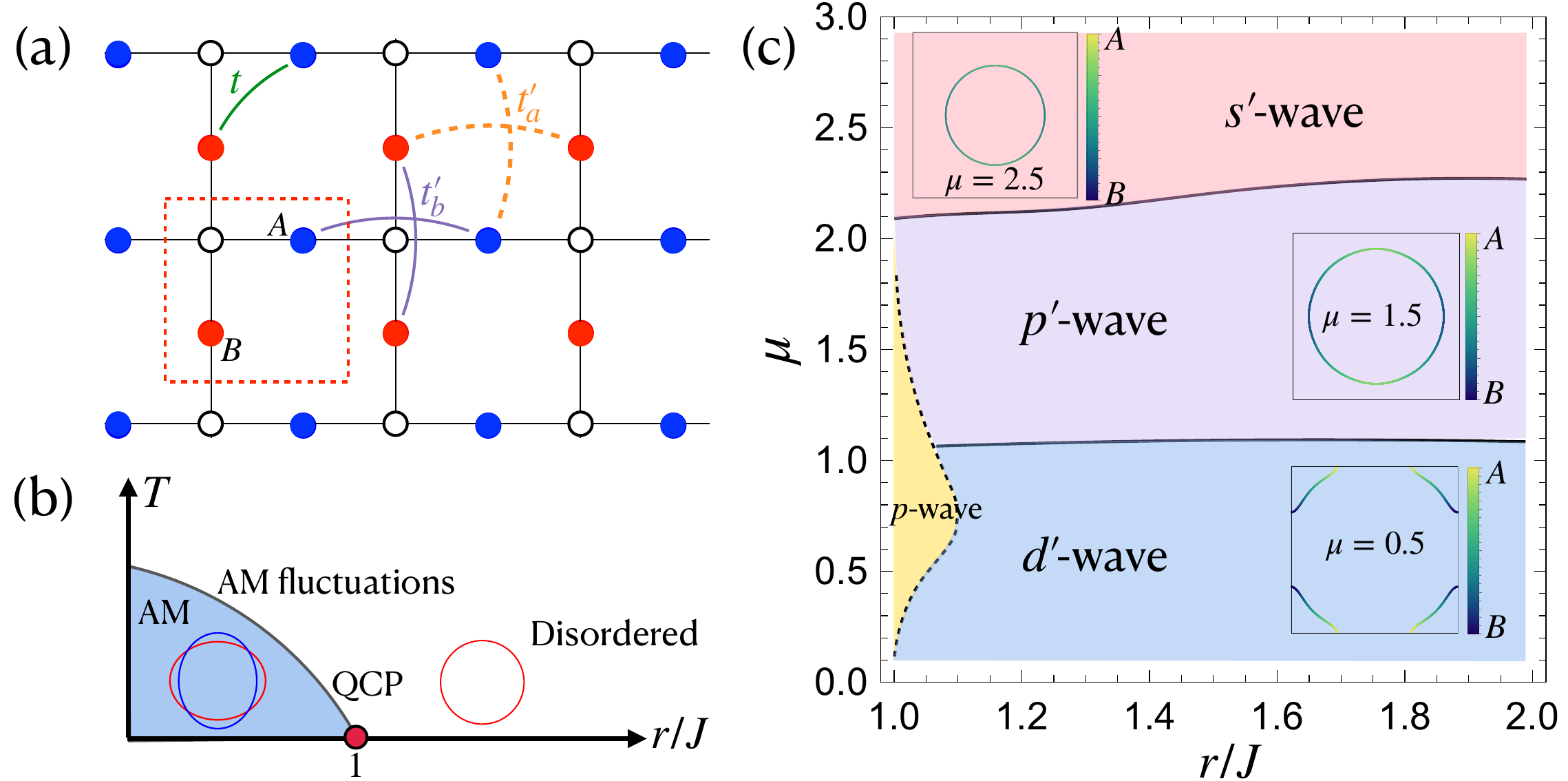}
    \caption{(a) Minimal AM model on the Lieb lattice \cite{antonenko2024mirrorchernbandsweyl}. The unit cell has two atomic positions ($A$ and $B$) related by a $90^\circ$ rotation, where up spins (red) and down spins (blue) form a spatially staggered configuration. (b) Schematic phase diagram of a putative AM QCP with tuning parameter $r$. Fluctuations are large when $r \gtrsim J$. (c)  Phase diagram of the leading pairing instabilities in the chemical potential ($\mu$) - $r$ plane. The corresponding pairing configurations are shown in Fig. \ref{fig:phase}, and representative Fermi surfaces for each $\mu$  range are shown in the insets where the color  gives the sublattice-projected spectral weight. The parameters used were $t=1$, $t_a'=-0.3$, and $t'_b=0.2$. 
}\label{fig:summary}
\end{figure}

The recent discovery of altermagnetism (AM)~\cite{Smejkal2020crystal,Smejkal2022PRX,JungwirthPRX2022} turned the spotlight on intra-unit-cell staggered spin orders, which, in contrast to the standard antiferromagnetic order, preserve translational symmetry. 
\YMW{A prototypical AM state defined on the Lieb lattice is shown in Fig.~\ref{fig:summary}(a), where anti-parallel spins are located on two atomic positions within the same unit cell that are related by a rotation~\cite{antonenko2024mirrorchernbandsweyl}. It breaks time-reversal symmetry and crystalline rotation symmetry, but preserves the product of the two.} As a result, in momentum space, the AM order parameter can be represented as a $\bm k$-dependent magnetization with a $d$-wave (or higher even-parity angular momentum) structure \cite{Jungwirth2024supefluid}. Interesting phenomena enabled by AM have been widely investigated, from topological and correlated phenomena \cite{JungwirthPRX2022,Smejkal_chiral_magnons,Bhowal2024,Fernandes2024_AM,Mcclarty2024,Leeb2024,AgterbergPRB2024,antonenko2024mirrorchernbandsweyl,Schnyder2024,Attias2024,Cano2024} to spintronic applications \cite{BaiPRL2023,Gonzalez-HernandezPRL2021,ZhangPRL2024,ChiPRApplied2024,SunPRB2023,HodtPRB2024,SmejkalPRX2022} to  unique effects emerging in superconducting heterostructures \cite{Ouassou2023,Neupert2023,Sun2023,Papaj2023,Wei2023,Beenakker2023,Li2023_Majorana,Li2024higherorder,Ghorashi2024}.

While the properties of a superconducting state emerging inside an altermagnetic state have been intensively studied \cite{Fradkin2014,Zhu2023,BanerjeePRB2024,ChakrabortyPRB2024,chakraborty2024perfectsuperconductingdiodeeffect,ParamekantiPRB2024,Scheurer2024,Sim2024pair,Carvalho2024,Heung2024,Hong2025unconventional}, pairing mediated by AM fluctuations has received less attention \cite{Mazin2022notes,Brekke2023,Maeland2024}. As a vast set of materials have been proposed or shown to be altermagnets \cite{Smejkal2022PRX,Facio2023,Sodequist2024,Haule2024}, it is important to elucidate this possibility. Since AM order is uniform and time-reversal-odd, it may be tempting to conclude that pairing should occur in the spin-triplet channel. Indeed, at long wavelengths, AM fluctuations behave just like Pomeranchuk fluctuations in the even-parity spin-channel \cite{Wu2007}, which favor spin-triplet $p$-wave pairing \cite{palle2024superconductivity}. 
However, as explained above, AM order has an intra-unit-cell staggered spatial structure. Thus, at short wavelengths, AM fluctuations are reminiscent of antiferromagnetic N\'eel fluctuations, which are known to favor $d$-wave spin-singlet pairing \cite{Mazin2022notes}.

In this work, we show that this dichotomy between long-wavelength and short-wavelength AM fluctuations leads to a rich AM-mediated superconducting (SC) phase diagram, controlled not only by the shape of the Fermi surface but also by the strength of the AM fluctuations. Using the well-established minimal model for AM on the Lieb lattice \cite{Brekke2023,antonenko2024mirrorchernbandsweyl}, illustrated in Fig.~\ref{fig:summary}(a), we solve the linearized gap equations for electrons coupled to AM fluctuations, whose strength is determined by the proximity to a putative quantum critical point (QCP), as shown in Fig.~\ref{fig:summary}(b). For weak to moderate fluctuations, as displayed in the phase diagram of Fig.~\ref{fig:summary}(c), we find that superconductivity is dominated by intra-unit-cell pairings,
\YMW{which can be of $s'$-, $d'$- and $p'$-wave symmetries depending on the shape of Fermi surface (the prime is used to indicate that pairing is intra-unit-cell).
In these pairing states, Cooper pairs are formed by electrons from different sublattices and the gap function changes sign upon a shift by a primitive reciprocal lattice vector. }
For stronger fluctuations, we recover the standard $p$-wave triplet state expected from the long wavelength behavior of the AM fluctuations, without intra-unit-cell order. 

We further determine the properties of these SC states in the coexistence state with long-range AM order. We find that the $d'$ pairing state becomes $d'+is'$, and exhibits Bogoliubov Fermi surfaces. On the other hand, inside the AM phase,  the $p\pm ip$ pairing state coexisting with AM order exhibits  corner Majorana zero modes that reflect a higher-order topological state protected by the combined fourfold rotational and time-reversal symmetries.

{\it Microscopic Model.}-- We consider itinerant electrons on the Lieb lattice, as shown in Fig.~\ref{fig:summary}(a), with fermionic destruction operators on the two sublattices denoted by $A_{\uparrow/\downarrow,\bk}$ and $B_{\uparrow/\downarrow,\bk}$. In the basis $\psi_{\bk}=(A_{\uparrow,\bk}, A_{\downarrow,\bk},B_{\uparrow,\bk}, B_{\downarrow,\bk})^T$, the free-fermion tight-binding Hamiltonian is given by \cite{antonenko2024mirrorchernbandsweyl}:
\begin{equation}
\mathcal{H}_0(\bk)=h_0(\bk)\tau_0\sigma_0+h_1(\bk)\tau_1\sigma_0+
h_3(\bk)\tau_3\sigma_0,\label{eq:H0k}
\end{equation}
where $\tau_i$ and $\sigma_i$ are Pauli matrices acting on sublattice and spin subspaces, respectively, with $h_0(\bk)=-(t_a'+t_b')(\cos k_x +\cos k_y)-\mu$, $h_1(\bk)=-4t\cos\frac{k_x}{2}\cos\frac{k_y}{2}$, and $h_3(\bk)=-(t_a'-t_b')(\cos k_x -\cos k_y)$. 
Here, $t$ is the nearest neighbor hopping and $t_a'$ and $t_b'$ are the two next-nearest neighbor hoppings.  Eq.~\eqref{eq:H0k} features a quadratic band touching point at $M=(\pi,\pi)$~\cite{SunPRL2009} for $\mu_0=2(t_a'+t_b')$. 
\YMW{As $\mu$ increases from $\mu_0$, the Fermi surface evolves from a small pocket around M point to a pocket around $\Gamma=(0,0)$ via a Lifshitz transition [see insets of Fig.~\ref{fig:summary}(c)].
Since the two sublattices are spatially separated by a half lattice vector $(1/2,1/2)$, $\mathcal{H}_0$ is not periodic under a shift by the primitive reciprocal lattice vector $\bm{G}_1=(2\pi,0)$ or $\bm{G}_2=(0,2\pi)$.
Instead, it satisfies}
\be
\mathcal{H}_0(\bk + \bm{G}_{1,2}) = \tau_3 \mathcal{H}_0(\bk) \tau_3.
\label{eq:nonsymm}
\ee
 
The AM order parameter in this model corresponds to the intra-unit-cell staggered magnetization $\boldsymbol{N}$  between sublattices $A$ and $B$. It couples to the electrons via $\mathcal{H}_\mathrm{AM}= N\tau_3\sigma_3$, thus breaking time-reversal ($\mathcal{T}$) and fourfold rotational symmetry ($C_4$), but preserving $C_4\mathcal{T}$. Here, we chose $N$ to point along the $z$-axis since, in the presence of SOC, this is the only moment direction for which the system remains a pure altermagnet, in the sense that no secondary weak ferromagnetic component is induced \cite{antonenko2024mirrorchernbandsweyl}. We will first analyze the problem without SOC and then discuss its role later. Although here we focus on the Lieb lattice, Hamiltonians of the same form as Eq.  \eqref{eq:H0k} describe a wide-range of AM low-energy models on different types of lattices \cite{AgterbergPRB2024}.

We consider superconducting instabilities from pairing interaction mediated by AM fluctuations in the disordered phase near a putative QCP, as illustrated in Fig.~\ref{fig:summary}(b). This is different from previous works on a related Lieb model, which focused on pairing by the double exchange of magnons in the AM ordered state \cite{Brekke2023,Maeland2024}. 
By symmetry, the form of the AM susceptibility $\chi_{\bq}$ is determined by a Hamiltonian similar to Eq.~\eqref{eq:H0k}, with hopping parameters replaced by spin-spin interactions. For our purposes, it is sufficient to consider only the dominant inter-sublattice interaction, which we denote by $J$, resulting in  $\chi_{\bq}^{-1}=r-J \cos\frac{q_x}{2}\cos\frac{q_y}{2}$, where $r$ measures the distance to the QCP. Clearly, the QCP is reached when $r/J=1$, signaled by the divergence of $\chi_{\bq=0}$. Promoting operators to fermionic and bosonic quantum fields, we obtain the action $S=S_0+S_{\mathrm{int}}$, with:

\begin{equation}
    \begin{aligned}
        &S_0=-\int_k\psi^\dagger_{k}\hat G_0^{-1}(k)\psi_{k}+
        \int_q N_{q} D^{-1}(q)N_{-q},\\
&S_{\mathrm{int}}=g\int_{k,q}N_q\psi_{k}^\dagger(\tau_3\sigma_3)\psi_{k+q}.\\
    \end{aligned}\label{eq:Stot}
\end{equation}
Here, we defined $\int_k=\frac{T}{V}\sum_{\omega_n,\bk}$ and $k=(i\omega_n,\bk)$ where $\omega_n$ is the Matsubara frequency. The propagators are given by $\hat G_0^{-1}(k)=i\omega_n-\mathcal{H}_0(\bk)$, and  $ D^{-1}(q)=\chi_{\bq}^{-1} + |\Omega_n|/\gamma_q$. In the last expression, $\gamma_q$ is the momentum-dependent Landau damping~\cite{Rech2006}, which can be calculated directly from the particle-hole bubble.

{\it SC from AM fluctuations.}-- Starting from Eq.~\eqref{eq:Stot}, we now integrate out the AM fluctuating field $N_q$  to obtain the AM-fluctuation mediated interaction between the electrons

\begin{equation}
    \tilde{S}_\text{int}=-g^2\int_{k,p,q} 
    D(q) \,
    \psi_{k}^\dagger(\tau_3\sigma_3)\psi_{k+q}\psi_{p}^\dagger(\tau_3\sigma_3)\psi_{p-q}.
    \label{eq:AMfluctuation}
\end{equation}

While the dynamics of the interaction is crucial for superconductivitity in the quantum critical regime~\cite{WuPRB2019,gamma1,gamma2,gamma3,gamma4}, for the purposes of identifying the leading pairing channels it suffices to consider its static component by setting $\Omega_n\to0$, yielding $D(q) \to \chi_{\bq}$. 
\YMW{Using the Fierz identities~\cite{VafekPRL2014,SavaryPRB2017,BoettcherPRB2016,WangPRB2021,WuPRB2023} related to the 16 matrices $\tau_i \sigma_j$ spanning the spin-sublattice space,
we can rewrite Eq.~\eqref{eq:AMfluctuation} in the Cooper channel (see the Supplementary Material, SM~\cite{SM}).
We find that only eight channels $\tau_i\sigma_j$ have attractive pairing interactions. Keeping only these channels results in:}
\begin{equation}
    S_{C}=-\sum_{i=1}^{8}\int_{k,p}\left(\psi^\dagger_{k}\Gamma^i \psi^*_{-k}\right)\tilde{\chi}_{\bk,\bp}\left(\psi^\text{T}_{-p}\Gamma^i\psi_{p}\right)\label{eq:SIC}
\end{equation}
where we defined $\tilde{\chi}_{\bk,\bp}\equiv\frac{g^2}{4} \chi_{\bk - \bp}$. Moreover, without SOC, only the following six channels give a nonzero $T_c$,
\begin{equation}
    \begin{aligned}
        &\Gamma^1=\tau_0\sigma_0, ~~ \Gamma^2=\tau_3\sigma_3, ~~\Gamma^3=\tau_3\sigma_0, ~~ \Gamma^4=\tau_0\sigma_3, \\
         &\Gamma^5=\tau_1\sigma_2, ~~ \Gamma^7=\tau_1\sigma_1, 
    \end{aligned}\label{eq:Gamma8}
\end{equation}
while the other two, $\Gamma^6=\tau_2\sigma_1$ and $\Gamma^8=\tau_2\sigma_2$, do not gap out the Fermi surface.

It is straightforward to derive the linearized gap equations from Eq.~\eqref{eq:SIC} by introducing the gap function $ \hat\Delta(\bk) =\sum_i\Delta_i(\bm k)\Gamma^i \sim \sum_i\psi_{\bk}^\dagger\Gamma^i  \psi^*_{-\bk}$ and integrating out the fermions:
\begin{equation}
    \Delta_i(\bp)=\sum_{\bk,j}\tilde{\chi}_{\bp,\bk}P^{i,j}_{\bk}\Delta_j(\bk),\label{eq:lineargap}
\end{equation}
where $P^{i,j}_{\bk}:=-T\sum_{\omega_n}\text{Tr}[\hat G_h(k)\Gamma^i\hat G_p(k)\Gamma^j]$; here, $\hat G_p(k)=i\omega_n-\mathcal{H}_0(\bk)$ and $\hat G_h(k)=i\omega_n+\mathcal{H}^T_0(-\bk)$ are the bare particle and hole Green's functions.

\begin{figure}
    \includegraphics[width=8.5cm]{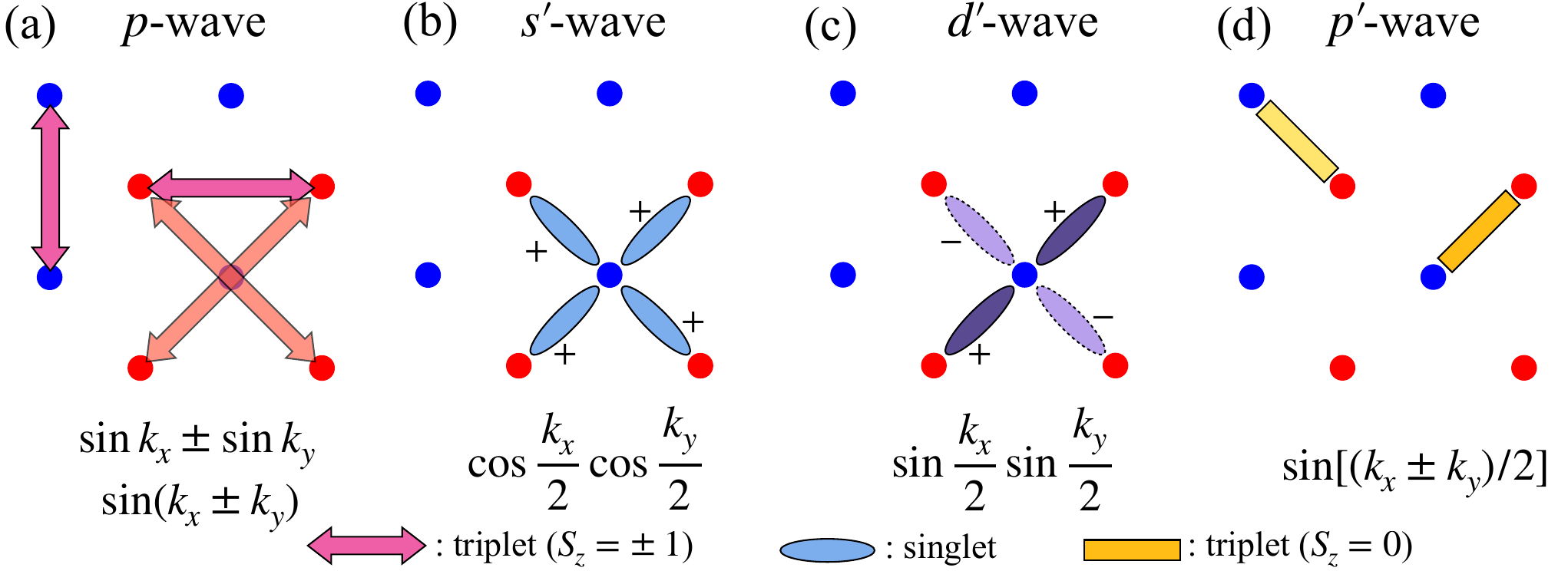}
    \caption{Real space configuration of the distinct pairing channels shown in Fig.~\ref{fig:summary}(c). The primes denote intra-unit-cell pairings  whose Cooper pairs are formed by electrons from different sublattices.}\label{fig:phase}
\end{figure}

\textit{Superconducting phase diagram.}--   By solving the gap equations, Eq.~\eqref{eq:lineargap}, we obtain the superconducting phase diagram in Fig. \ref{fig:summary}(c) for  $t=1$,  $ t_a'=-0.3$ and $t_b'=0.2$,  as a function of $r/J$, the distance to the QCP, and $\mu$. The details of the method, the symmetry of the pairing orders, and the analysis of a wider parameter regime are discussed in the SM~\cite{SM}. Note that, as this is a weak-coupling approach, the results as the QCP is approached ($r/J\to1$) should be understood as representing the tendency of which pairing channel is selected, rather than an accurate calculation of $T_c$.  

We first discuss the $p$-wave state, which has fourfold degeneracy and has nonzero components $\Delta_i(\bk)$ from combinations of the $\Gamma^1,\Gamma^2,\Gamma^3,\Gamma^4$ channels. The four degenerate gap functions can be written as:

\begin{align}
        &\hat \Delta_{(1)}(\bk)=\Delta_{(1)}\sigma_{\uparrow}[\tau_0v_+(\bk)+\tau_3v_-(\bk)],\nonumber\\
         &\hat \Delta_{(2)}(\bk)=\Delta_{(2)}\sigma_{\uparrow}[\tau_0v_-(\bk)+\tau_3v_+(\bk)],\nonumber\\
         &\hat \Delta_{(3)}(\bk)=\Delta_{(3)}\sigma_{\downarrow}[\tau_0v_+(\bk)+\tau_3v_-(\bk)],\nonumber\\
         &\hat \Delta_{(4)}(\bk)=\Delta_{(4)}\sigma_{\downarrow}[\tau_0v_-(\tilde \bk)+\tau_3v_+(\tilde \bk)],
\label{eq:fourcomponents}
\end{align}
where $\sigma_{\uparrow/\downarrow} = (\sigma_0\pm\sigma_3)/2$ is the spin projection. This form makes it clear that these are fourfold-degenerate $S_z=\pm1$ triplet pairing states involving electrons from the same sublattice $A$ or $B$. The odd-parity functions $v_\pm(\bk)$ are obtained by solving the gap equations Eq.~\eqref{eq:lineargap}. As we show in the SM~\cite{SM},  an expansion to the leading lattice harmonics gives$v_\pm(\bk)\approx (\sin k_x\pm \sin k_y)-\sin(k_x\pm k_y)$, corresponding thus to $p$-wave gap functions. Their real-space patterns are illustrated in Fig.~\ref{fig:phase}(a), corresponding to pairing between same-sublattice nearest neighbors. 

We now proceed to the $s'$- and $d'$-wave parings, both of which have a nonzero gap component $\Delta_5(\bk)$ only in the $\Gamma^5 = \tau_1\sigma_2$ channel. Since $\Gamma^5$ is antisymmetric and off-diagonal in sublattice space, it corresponds to spin-singlet and intra-unit-cell pairing of the form $ A^{\dagger}_{\uparrow,\bk}B^{\dagger}_{\downarrow,-\bk}$,  involving two electrons from different sublattices. Thus, similar to Eq.~\eqref{eq:nonsymm}, the even-parity gap function $\Delta_5(\bk)$ must obey $\Delta_5(\bk+\bm{G}_{1,2})=-\Delta_5(\bk)$. As a result, the gap functions cannot have either the standard $s$-wave form factor (a constant) or the standard $d_{xy}$-wave form factor ($\sin k_x \sin k_y$). Instead, as we verify by directly solving the gap equations (see the SM \cite{SM}), the dominant lattice harmonics are $\Delta_5(\bk)=\Delta_s \cos\frac{k_x}{2}\cos\frac{k_y}{2}$ and  $\Delta_5(\bk)=\Delta_d \sin\frac{k_x}{2}\sin\frac{k_y}{2}$, which we denote as $s'$-wave and $d'$-wave states, respectively (the primes indicate intra-unit-cell pairing). The real-space structure of these gaps are illustrated in Fig.~\ref{fig:phase}(c,d). 

\begin{figure}
    \includegraphics[width=8.cm]{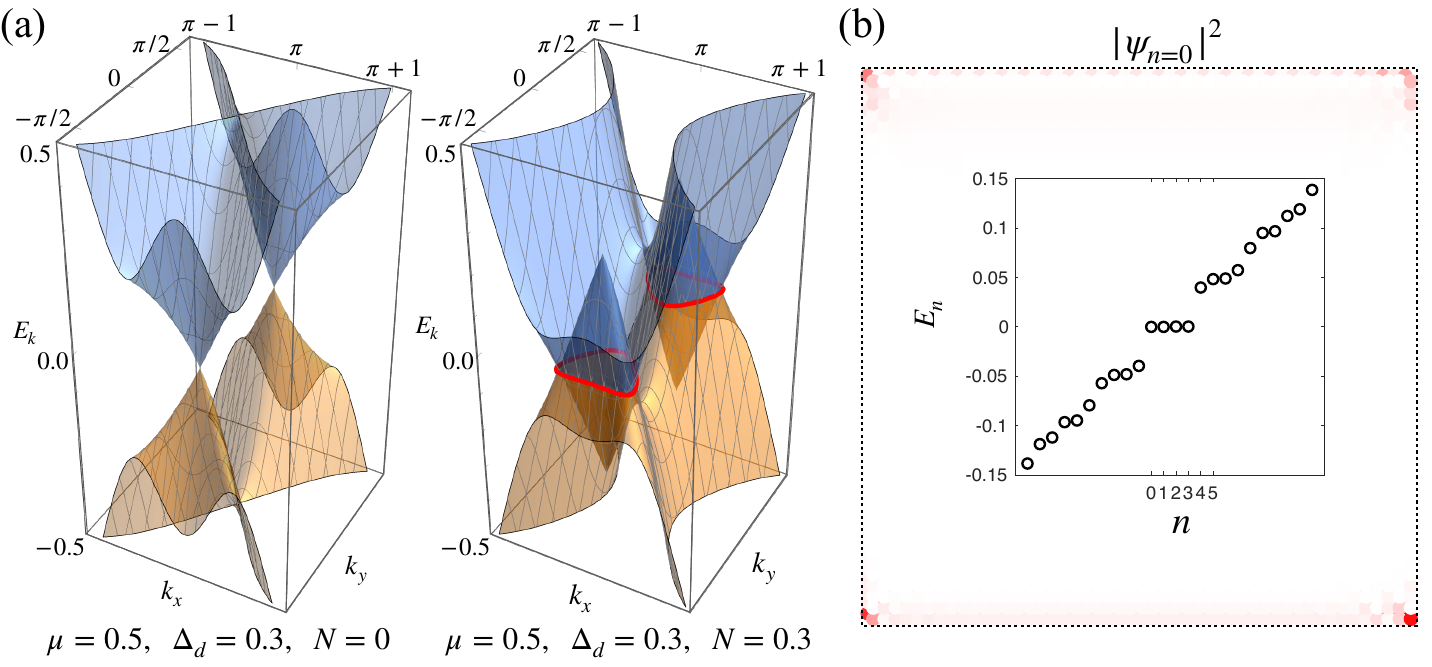}
    \caption{(a) BdG spectrum of the $d'$-state at the BZ boundary, exhibiting  Dirac nodes (without AM order, left) and Bogoliubov Fermi surfaces (with AM order, right). (b) Eigenvalues (inset) and the $n=0$ wavefunction obtained by diagonalizing the BdG Hamiltonian of the $p_B$-phase in the presence of AM order $N=0.3$, showcasing the  Majorana corner modes. Here, $n$ labels the eigenvalues of the finite-size system.}
    \label{fig:corner}
\end{figure}

The $s'$-wave function has zeroes along the Brillouin zone (BZ) boundaries $k_x = \pm \pi$ and  $k_y = \pm \pi$, and the $d'$-wave function has zeros along $k_x=0$ and $k_y=0$. Importantly, additional zeroes emerge from the projection of the intra-unit-cell pairing vertex $\Gamma^5= \tau_1\sigma_2$ onto the Fermi surface. Indeed, because of $\tau_1$, the projected gap must vanish whenever the associated Fermi surface state has pure $A$ or $B$ character. According to Eq.~\eqref{eq:H0k}, this is always the case along the BZ boundaries. Thus, depending on the characteristics of the underlying Fermi surface, the projected gap function of the $s'$-wave state has quadratic nodes at the BZ boundary, and the $d'$-wave state acquire sign-changing nodes, corresponding to Dirac points in the BdG spectrum [see Fig.~\ref{fig:corner}(a)]. As we show in the SM \cite{SM}, the Dirac points in the $d'$-wave state are enforced by symmetry. 

Finally, the $p'$-wave pairing has nonzero components only in the  $\Gamma^7 = \tau_1\sigma_1$ channel, which is spin-triplet with $S_z=0$ but also intra-unit-cell. Although both $p$-wave and $p'$-wave states transform as the same irreducible representation of the tetragonal group, they have different transformation properties in spin space.  As such, the gap functions are distinct from the $p$-wave state in Eq. (\ref{eq:fourcomponents}). Instead, by solving the gap equations, we find the degenerate gaps $\hat\Delta_{(1)}(\bk)=\Delta_{(1)}\Gamma^7u_+(\bk)$ and $\hat\Delta_{(2)}(\bk)=\Delta_{(2)}\Gamma^7u_-(\bk)$. The form factors are dominated by the lattice harmonics  $u_\pm(\bk)=\sin[(k_x \pm k_y)/2] $, which are compatible with a sign change under a shift by $\bm{G}_1$ or $\bm{G}_2$. The real-space patterns of this $p'$-wave state are presented in Fig.~\ref{fig:phase} (d). 

\YMW{The regimes where each of these four pairings ($p$-, $s'$-, $d'$-, and $p'$-wave) is the leading superconducting instability is shown in Fig. \ref{fig:summary}(c), which can be understood qualitatively. First, intra-unit-cell pairing is generally favored farther away from the putative QCP, where the AM correlation length is shorter and hence the staggered spin structure of the AM state is most prominently manifested. In contrast, when the correlation length is long enough, the AM fluctuations behave like the excitations above a spin-triplet Pomeranchuk instability, which favor a standard $p$-wave state.
The leading intra-unit-cell pairing in the regime of shorter-range fluctuations can be understood from the evolution of the Fermi surface as a function of $\mu-\mu_0$ [see the insets of Fig. \ref{fig:summary}(c)]. For small $\mu-\mu_0$, the Fermi surface is a small pocket around the $M$ point. Since the magnitude of the $d'$-wave form factor $|\sin\frac{k_x}{2}\sin\frac{k_y}{2}|$ is the largest at $M$, it is reasonable that the leading pairing is the $d'$-wave in this case. 
For large $\mu-\mu_0$, the Fermi surface is small but centered at the $\Gamma=(0,0)$ point. In this case, it is the form factor of the $s'$-wave that has the largest magnitude. For intermediate values of $\mu-\mu_0$, the Fermi surface is large and remains close to the BZ boundary, disfavoring both the $s'$-wave and $d'$-wave states, which have nodes either on the BZ boundaries or along $k_{x,y}=0$. As a result, the $p'$-wave becomes the leading instability.} 

\textit{Topological superconductivity}-- The degeneracy within  the multi-component $p$-wave (fourfold) and $p'$-wave (twofold) states can result in various topological superconducting states.
In the case of the $p'$-wave state, we find that the quartic terms of the Ginzburg-Landau (GL) free energy expansion are minimized when the two components $\Delta_{(1)}$ and $\Delta_{(2)}$ form a $p'\pm ip'$ chiral SC. As we show in the SM \cite{SM}, the corresponding BdG Hamiltonian belongs to class-D topological superconductivity \cite{topoclassification}. The BdG spectrum is then fully gapped, each band is twofold degenerate, and the total Chern number $C$ for the bands below the Fermi level is $C=2$.

As for the $p$-wave state described by Eq.~\eqref{eq:fourcomponents}, the GL expansion reveals that the gaps associated with the spin-up projection $\left( \Delta_{(1)},\,\Delta_{(2)}\right)$ form a $(p\pm ip)_\uparrow$ condensate whereas the spin-down projection gaps $\left( \Delta_{(3)},\,\Delta_{(4)}\right)$ form a $(p\pm ip)_\downarrow$ condensate. Depending on the relative chirality of each condensate, the result is either a time-reversal symmetry-breaking phase similar to the $A$-phase of superfluid  $^3$He, which we denote $p_A$-phase (same chirality), or a time-reversal symmetry-preserving phase similar to the $B$-phase, denoted $p_B$-phase (opposite chirality). Both are topologically nontrivial: the $p_A$ phase has $C=2$, and the $p_B$-phase corresponds to a topological SC in class DIII with a nontrivial $\mathbb{Z}_2$ index, which exhibits helical edge states\cite{topoclassification}.

\YMW{When coexisting with the AM order (which is possible since AM order produces no uniform magnetization), some of the SC orders may also display nontrivial topological characteristics.
For the $d'$-wave state, the presence of AM order transforms the Dirac nodes at the BZ boundary into Bogoliubov Fermi surfaces (BFSs), as shown in Fig.~\ref{fig:corner}(a). These BFSs are captured by a nontrivial $\mathbb{Z}_2$ topological invariant, which is protected by particle-hole  and inversion symmetries \cite{AgterbergPRL2017, BzdusekPRB2017, santos-wang-fradkin}. }Furthermore, as we show in the SM \cite{SM}, the coexistence between the $d'$-wave state and the AM order parameter $N$ induces a separate $s'$-wave order parameter $\tilde\Delta_s$ (with the same symmetry as $\Delta_s$) via a trilinear coupling term $N\Imm(\tilde\Delta_s\Delta_d^*)$ in the GL free energy. The corresponding pairing state in general becomes $d'+is'$ inside the AM phase. However, as no additional symmetry is broken by the subdominant $\tilde\Delta_s$ component, the BFSs remains robust.

For $p$-wave pairing, the time-reversal symmetry of the $p_B$-phase is broken when in coexistence with AM. As a result, the nontrivial $\mathbb{Z}_2$ topology is eliminated, and edges are gapped. However, since the AM order changes sign upon a $C_4$ rotation, it vanishes along the $\hat e_x\pm \hat e_y$ directions. Thus, by analogy with the Jackiw-Rebbi model~\cite{JackiwRebbiPRD1976}, at each corner of the sample there exists a single Majorana zero mode, resulting in a higher-order topological state~\cite{WangLinHughes2018,JahinPRR2021}. The corner modes are shown numerically in Fig.~\ref{fig:corner}(b) by diagonalizing the (real space) BdG Hamiltonian in a $49\times49$ system with open boundary conditions. A similar phenomenon was also obtained in Ref.~\cite{ZhuPRB2023} in a different context where SC is induced in an AM via the proximity effect.

\textit{Effects of the SOC}-- While altermagnetism is formally defined in the absence of SOC, it is legitimate to consider how SOC affects AM-related phenomena. In the Lieb lattice model of Eq.~\eqref{eq:H0k}, a Kane-Mele SOC term of the form $h_2(\bk)\tau_2\sigma_3=\lambda\sin\frac{k_x}{2}\sin\frac{k_y}{2}\tau_2\sigma_3$ is allowed~\cite{antonenko2024mirrorchernbandsweyl}. 
The presence of SOC introduces small mixing between $\Gamma^5$ and $\Gamma^6$ for the $s'$- and $d'$-wave pairings, and between $\Gamma^7$ and $\Gamma^8$ for the $p'$-wave pairing. However, the weights from $\Delta_6(\bk)$ and $\Delta_8(\bk)$ remains small even with $\lambda=0.3$. Moreover, this type of SOC lowers the four-fold degeneracy of the $p$-wave pairing into $2\times$two-fold degeneracy by mixing spins and sublattices. The leading pairing phase diagram remains largely unchanged, as shown in the SM \cite{SM}. 

\textit{Conclusion.}-- In this letter, we showed that, because of the existence of sublattice degrees of freedom, AM fluctuations mediate not only standard $p$-wave superconductivity, but also intra-unit-cell pairing. The latter can have triplet $p'$-wave, singlet $s'$-wave, or singlet $d'$-wave character depending on the Fermi surface characteristics. Interestingly, different topological superconducting states emerge in both the AM-disordered and AM-coexistence states. Suppressing the AM transition via non-thermal tuning parameters, particularly in metallic altermagnets, is a promising path towards this novel scenario for unconventional superconductivity. The resulting intra-unit-cell pairing states, whose gap functions change sign when the momentum is shifted by a primitive reciprocal lattice vector, should be visible by local probes such as scanning tunneling microscopy. More broadly, our work reveals a rich landscape of unconventional and topological superconductivity near the onset of AM order.

\begin{acknowledgments}
We would like to thank D.~Agterberg, D. Antonenko, E. Fradkin, M.~Franz, J.~Hamlin, P.~Hirschfeld, I.~Mazin, J. Venderbos, Hong Yao, Zhengzhi Wu, and X. Zou for useful discussions. Y.-M.W. acknowledges support from the Gordon and Betty Moore Foundation’s
EPiQS Initiative through GBMF8686. YW is supported by NSF under award number DMR-2045781. R.M.F. was supported by the Air Force Office of Scientific Research under Award No. FA9550-21-1-0423. 
\end{acknowledgments}

\bibliography{AlterM}

\end{document}


\onecolumngrid
  \begin{center}
  \textbf{  SUPPLEMENTARY MATERIAL FOR
  \\[.2cm]\large Intra-unit-cell singlet pairing mediated by altermagnetic fluctuations}
\\[0.2cm]
Yi-Ming Wu,$^{1}$, Yuxuan Wang$^{2}$ and Rafael Fernandes$^{3,4}$
\\[0.2cm]
{\small \it $^{1}$ Stanford Institute for Theoretical Physics, Stanford University, Stanford, California 94305, USA}

{\small \it $^{2}$ Department of Physics, University of Florida, Gainesville, Florida 32611, USA}

{\small \it $^{3}$ Department of Physics, The Grainger College of Engineering, University of Illinois Urbana-Champaign, Urbana, Illinois 61801, USA}

{\small \it $^{4}$ Anthony J. Leggett Institute for Condensed Matter Theory, The Grainger College of Engineering, University of Illinois Urbana-Champaign, Urbana, Illinois 61801, USA}

\end{center}

\setcounter{equation}{0}
\renewcommand{\theequation}{S\arabic{equation}}
\setcounter{figure}{0}
\renewcommand{\thefigure}{S\arabic{figure}}
\renewcommand{\thetable}{S\Roman{table}}
\setcounter{page}{1}

\tableofcontents


\section{I. Fierz identity} 
\label{sec:i_diagrammatic_representation_of_fiertz_identity}
In this section, we present a simple diagrammatic derivation of Fierz identity, and then apply it to obtain the Cooper channel  action, Eq. (5) of the main text, from the altermagnetic fluctuation induced interaction, Eq. (4) of the main text. 

\subsection{1. Diagrammatic representation of Fierz identity} 
\label{sub:1_diagrammatic_representation_of_fierz_identity}

Let $\Gamma^a, a=1,...,n^2$ be the set of orthonormal basis that span the space of $n\times n$ Hermitian matrices. The case of $n=4$ is the main focus of our study. The corresponding $16$ matrices are given in terms of the spin and sublattice Pauli matrices as:
\begin{equation}
    \begin{aligned}
        &\Gamma^1=\tau_0\sigma_0, ~~ \Gamma^2=\tau_3\sigma_3, ~~\Gamma^3=\tau_3\sigma_0, ~~ \Gamma^4=\tau_0\sigma_3,\\
        &\Gamma^5=\tau_1\sigma_2, ~~ \Gamma^6=\tau_2\sigma_1, ~~\Gamma^7=\tau_1\sigma_1, ~~ \Gamma^8=\tau_2\sigma_2, ~ \\
        &\Gamma^9=\tau_2\sigma_0, ~~ \Gamma^{10}=\tau_2\sigma_3, ~~\Gamma^{11}=\tau_1\sigma_0, ~~ \Gamma^{12}=\tau_1\sigma_3,  \\
        &\Gamma^{13}=\tau_0\sigma_1, ~~ \Gamma^{14}=\tau_0\sigma_2, ~~\Gamma^{15}=\tau_3\sigma_1, ~~ \Gamma^{16}=\tau_3\sigma_2
    \end{aligned}
\end{equation}

The completeness of this basis is expressed via
\begin{equation}
    \frac{1}{n}\sum_{a}\Gamma^a_{ij}\Gamma^a_{lm}=\delta_{im}\delta_{jl}\label{eq:completeness}
\end{equation}
such that any Hermitian matrix $M$ in this space can be uniquely written as 
\begin{equation}
    M=\frac{1}{n}\sum_a \text{Tr}(M\Gamma^a)\Gamma^a.
\end{equation}
In all the summations, $a$ runs from $1$ to $n^2$.
To develop a diagrammatic representation of the Fierz identities, we  rewrite Eq.~\eqref{eq:completeness} diagrammatically as
\begin{equation}
    \begin{tikzpicture}[
  vertex/.style = {circle, draw, inner sep=1pt, fill=white},
  vertex1/.style = {vertex, fill=red!30!white},
  vertex2/.style = {vertex, fill=orange!30!white},
  vertex3/.style = {vertex, fill=blue!30!white},
  vertex4/.style = {vertex, fill=teal!30!white},
]
  \draw[thick, dashed]  (0,2)  -- (0,1) node[right] {$a$}-- (0,0) ;
  \draw[thick,postaction={on each segment={mid arrow=black}}] (-2,2)node[above]{$l$}  -- (0,2) node[vertex3] {$\Gamma$}  -- (2,2) node[above]{$m$};
   \draw[thick, postaction={on each segment={mid arrow=black}}] (-2,0)node[above]{$i$} -- (0,0) node[vertex3] {$\Gamma$} -- (2,0)node[above]{$j$};
   \draw (3,1)node {$=$};
   \draw[thick,postaction={on each segment={mid arrow=black}}] (4,2) node[above] {$l$} -- (5,1)--(6,0)node[above] {$j$} ;
   \draw[thick,postaction={on each segment={mid arrow=black}}] (4,0)node[above] {$i$} -- (4.9,0.9)
   (5.1,1.1)--(6,2)node[above] {$m$};

\end{tikzpicture}\label{eq:complete}
\end{equation}
The rule here is that the arrows at each vertex $i,j,m,l$ must be preserved, e.g. the arrow at $i$ points inward while the arrow at $j$ points outward.  
A matrix element is denoted by
\begin{equation}
    \begin{tikzpicture}[
  vertex/.style = {circle, draw, inner sep=1pt, fill=white},
  vertex1/.style = {vertex, fill=red!30!white},
  vertex2/.style = {vertex, fill=orange!30!white},
  vertex3/.style = {vertex, fill=blue!30!white},
  vertex4/.style = {vertex, fill=teal!30!white},
]
    \draw (-1.8,0) node {$M_{ij}=$};
    \draw[thick,postaction={on each segment={mid arrow=black}}] (-1,0)node[above]{$i$}--(0,0)node[vertex2]{$M$}-- (1,0)node[above]{$j$};       
    \draw (1.5,0) node {$=$};
    \draw[thick,postaction={on each segment={mid arrow=black}}] (2,0)node[above]{$j$}--(3.5,0)node[vertex2]{$M^t$}-- (5,0)node[above]{$i$};
    \draw (5.8,0) node {$=M^T_{ji}.$};
    \end{tikzpicture}
\end{equation}
where $M^T$ is the transpose of $M$.

Usually, a four-fermion interaction term has the form $(\psi^\dagger M\psi) (\psi^\dagger N\psi)=\sum_{i,j,m,n}\psi^\dagger_i\psi_j\psi^\dagger_m\psi_nM_{ij}N_{mn}$ (the matrix $N$ here should not be confused with the AM order parameter in the main text). One then needs to recombine the fermions by expressing $M_{ij}N_{mn}$ in terms of the basis matrices $\Gamma^a$.  Diagrammatically, the product of two matrix elements can be manipulated in the particle-hole channel as 
\begin{equation}
    \begin{aligned}
        &\begin{tikzpicture}[
  vertex/.style = {circle, draw, inner sep=1pt, fill=white},
  vertex1/.style = {vertex, fill=red!30!white},
  vertex2/.style = {vertex, fill=orange!30!white},
  vertex3/.style = {vertex, fill=blue!30!white},
  vertex4/.style = {vertex, fill=teal!30!white},
]
    \draw (0,-0.5) node {$M_{ij}N_{mn}~=$};
    \draw[thick,postaction={on each segment={mid arrow=black}}] (1.5,0)node[above]{$i$}--(3,0)node[vertex2]{$M$}-- (4.5,0)node[above]{$j$};
    \draw[thick,postaction={on each segment={mid arrow=black}}] (4.5,-1)node[above]{$m$}--(3,-1)node[vertex2]{$N$}-- (1.5,-1)node[above]{$n$} (5,-0.5) node {$=$};
    \draw[thick,postaction={on each segment={mid arrow=black}}] (5.5,0)node[above] {$i$}.. controls (6.2,-1)  .. (7.,-1) node[vertex2] {$M$} (7,-1).. controls (7.8,-1)  .. (8.5,0)node[above] {$j$};
    \draw[very thick, white] (5.9,-0.55) -- (5.82,-0.45)  (8.11,-0.55) -- (8.18,-0.45);
    \draw[thick,postaction={on each segment={mid arrow=black}}] (8.5,-1) node[below] {$m$}.. controls (7.8,0)  .. (7.,0) node[vertex2] {$N$} (7,0).. controls (6.2,0)  .. (5.5,-1)node[below] {$n$} (9.2,-0.5)node {$=$};
    \draw[thick,dashed] (10,-0.5)--(10.5,-0.5) node[above]{$a$}-- (11,-0.5) (13,-0.5)--(13.5,-0.5)node[above]{$b$}--(14,-0.5) ;
    \draw[thick,postaction={on each segment={mid arrow=black}}] (10,0.5) node[right]{$i$}-- (10,-0.5) node[vertex3] {$\Gamma$} -- (10, -1.5)node[right]{$n$};
    \draw[thick,postaction={on each segment={mid arrow=black}}] 
    (11,-0.5) node[vertex3] {$\Gamma$}
    .. controls(11.3,-1.15)..
    (12,-1.25) node[vertex2] {$M$}
    .. controls(12.7,-1.15)..
    (13,-0.5) node[vertex3] {$\Gamma$} 
    .. controls(12.7,0.15)..
    (12,0.25) node[vertex2] {$N$} 
    .. controls(11.3,0.15).. 
     (11,-0.5);
    \draw[thick,postaction={on each segment={mid arrow=black}}] (14,-1.5)node[left]{$m$}-- (14,-0.5) node[vertex3] {$\Gamma$} -- (14, 0.5)node[left]{$j$};
    \end{tikzpicture}\\
    &~~~~~~~~~~~~~=\frac{1}{n^2}\sum_{a,b}\text{Tr}\left(\Gamma^a M \Gamma^b N\right)\Gamma^a_{in}\Gamma^b_{mj}.
    \end{aligned}\label{eq:phchannel}
\end{equation}
We have used Eq.~\eqref{eq:complete} to obtain the loop structure in the first line of Eq.~\eqref{eq:phchannel}, which is nothing but the trace of four matrices. The order of the matrices  inside the trace is dictated by the arrow of the loop. 
Note that when combining these matrix elements with fermions, one should be careful about the anti-commutation rule for the fermion operators.

When written in the particle-particle channels, we have 
\begin{equation}
    \begin{aligned}
        &\begin{tikzpicture}[
  vertex/.style = {circle, draw, inner sep=1pt, fill=white},
  vertex1/.style = {vertex, fill=red!30!white},
  vertex2/.style = {vertex, fill=orange!30!white},
  vertex3/.style = {vertex, fill=blue!30!white},
  vertex4/.style = {vertex, fill=teal!30!white},
]
    \draw (0,-0.5) node {$M_{ij}N_{mn}~=$};
    \draw[thick,postaction={on each segment={mid arrow=black}}] (1.5,0)node[above]{$i$}--(3,0)node[vertex2]{$M$}-- (4.5,0)node[above]{$j$};
    \draw[thick,postaction={on each segment={mid arrow=black}}] (4.5,-1)node[above]{$m$}--(3,-1)node[vertex2]{$N$}-- (1.5,-1)node[above]{$n$} (5,-0.5) node {$=$};
    \draw[thick,postaction={on each segment={mid arrow=black}}] (5.5,0)node[above] {$i$}.. controls (6.2,-1)  .. (7.,-1) node[vertex2] {$M$} (7,-1).. controls (7.8,-1)  .. (8.5,0)node[above] {$j$};
    \draw[very thick, white] (5.9,-0.55) -- (5.82,-0.45)  (8.11,-0.55) -- (8.18,-0.45);
    \draw[thick,postaction={on each segment={mid arrow=black}}] (8.5,-1) node[below] {$n$}.. controls (7.8,0)  .. (7.,0) node[vertex2] {$N^t$} (7,0).. controls (6.2,0)  .. (5.5,-1)node[below] {$m$} (9.2,-0.5)node {$=$};
    \draw[thick,dashed] (10,-0.5)--(10.5,-0.5) node[above]{$a$}-- (11,-0.5) (13,-0.5)--(13.5,-0.5)node[above]{$b$}--(14,-0.5) ;
    \draw[thick,postaction={on each segment={mid arrow=black}}] (10,0.5) node[right]{$i$}-- (10,-0.5) node[vertex3] {$\Gamma$} -- (10, -1.5)node[right]{$m$};
    \draw[thick,postaction={on each segment={mid arrow=black}}] 
    (11,-0.5) node[vertex3] {$\Gamma$}
    .. controls(11.3,-1.15)..
    (12,-1.25) node[vertex2] {$M$}
    .. controls(12.7,-1.15)..
    (13,-0.5) node[vertex3] {$\Gamma$} 
    .. controls(12.7,0.15)..
    (12,0.25) node[vertex2] {$N^t$} 
    .. controls(11.3,0.15).. 
     (11,-0.5);
    \draw[thick,postaction={on each segment={mid arrow=black}}] (14,-1.5)node[left]{$n$}-- (14,-0.5) node[vertex3] {$\Gamma$} -- (14, 0.5)node[left]{$j$};
    \end{tikzpicture}\\
    &~~~~~~~~~~~~~=\frac{1}{n^2}\sum_{a,b}\text{Tr}\left(\Gamma^a M \Gamma^b N^T \right)\Gamma^a_{im}\Gamma^b_{nj}.
    \end{aligned}\label{eq:pp}
\end{equation}
Similar expression can be obtained using $M^T$ instead of $N^T$. 
The diagrammatic representation discussed here can be generalized to more complicated cases, and is helpful in keeping track of the matrix indices, as well as the multiplication order of matrices inside a trace.

\subsection{2. Application of Fierz identity in the Cooper channel} 
\label{sub:2_application_of_fierz_identity_in_the_cooper_channel}

We now derive the Cooper channel action using the Fierz identity demonstrated above. As noted in Eq. (4) of the main text, the interacting action is given by some density-density interaction
\begin{equation}
    \begin{aligned}
      \tilde{S}_\text{int}&=-g^2\int_{k,p,q} 
    \chi(\bq) \,
    \psi_{k}^\dagger M\psi_{k+q}\psi_{p}^\dagger N\psi_{p-q}\\
     &=-g^2\sum_{ijmn}\int_{k,p,q} 
    \chi(\bq) \,
    \psi_{i,k}^\dagger M_{ij}\psi_{j,k+q}\psi_{m,p}^\dagger N_{mn}\psi_{n,p-q}.
    \end{aligned}
\end{equation}
For the altermagnetic-fluctuation induced interaction, we have $M=N=\tau_3\sigma_3$. To study superconductivity, we need the Fierz identity in the particle-particle channel as in Eq. \eqref{eq:pp}. After shifting the momentum, we obtain
\begin{equation}
    \begin{aligned}
      \tilde{S}_\text{int}=-g^2\sum_{ijmn}\int_{k,p,q} 
    \chi(\bp-\bk) \frac{1}{16}\sum_{a,b}\text{Tr}(\Gamma^aM\Gamma^bN^t)
    \psi_{i,k}^\dagger \Gamma^a_{im}\psi_{m,p}^\dagger \psi_{n,p-q}\Gamma^b_{nj}\psi_{j,k+q}.
    \end{aligned}
\end{equation}
For $M=N=\tau_3\sigma_3$, if we treat $a,b=1,...,16$ as a two-matrix index, then
\begin{equation}
  \text{Tr}(\Gamma^aM\Gamma^bN^t)=4\times\text{diag}(1,1,1,1,1,1,1,1,-1,-1,-1,-1,-1,-1,-1,-1),
\end{equation}
i.e. it is diagonal in $a,b$ . The first eight elements correspond to an attractive pairing interaction and the last eight, to a repulsive pairing interaction. To search for uniform superconductivity, we thus keep only the first eight channels, and set the Cooper pair center of mass momentum frequency $q=0$. This leads to Eq. (5) of the main text,
\begin{equation}
    S_{C}=-\sum_{i=1}^{8}\int_{k,p}\left(\psi^\dagger_{k}\Gamma^i \psi^*_{-k}\right)\tilde{\chi}_{\bk,\bp}\left(\psi^\text{T}_{-p}\Gamma^i\psi_{p}\right)\label{eq:Cooperaction}
\end{equation}
where 
\begin{equation}
  \tilde{\chi}_{\bk,\bp}=\frac{g^2}{4}\chi(\bp-\bk)=\frac{g^2}{4}\chi(\bk-\bp), ~~~\chi(\bk-\bp)=\frac{1}{r-J(\bk-\bp)}, ~~J(\bq)=J\cos\frac{q_x}{2}\cos\frac{q_y}{2}.\label{eq:chirJ}
\end{equation}


\section{II. Symmetrized linearized gap equation} 
\label{sec:symmetrized_l}

\subsection{1. Derivation of the linearized gap equation} 
\label{sub:1_derivation_of_the_linearized_gap_equation}

To determine which pairing channel gives the leading pairing instability, we will derive the linearized gap equation and solve for the highest $T_c$. To this end, we introduce the auxiliary field $\Delta_{i}$ (a column vector) and decompose Eq.~\eqref{eq:Cooperaction} via the Hubbard-Stratonovich (HS) transformation
\begin{equation}
    S_{C}\to S=\sum_i\Delta_i^\dagger\tilde\chi^{-1}\Delta_i+\Delta_i^\dagger b_i+b_i^\dagger\Delta_i
\end{equation}
where $[\Delta_i]_{\bk}=\Delta_i(\bk)$ is the gap function and $[b_i]_{\bp}$ is the abbreviation for $\psi^\text{T}_{-\bp}\Gamma^i\psi_{\bp}$.
Integrating out the fermions results in the following action
\begin{equation}
    S[\Delta]=\Delta_i^\dagger\tilde\chi^{-1}\Delta_i-\text{Tr}\ln\left[\mathcal{G}_0^{-1}+\Sigma\right]\label{eq:tracelog}
\end{equation}
where 
\begin{equation}
    [\hat{\mathcal{G}}_0]_{k,p}=\delta_{k,p}\begin{pmatrix}
        \hat G_p(k) & 0\\
        0 & \hat G_h(k)
    \end{pmatrix}, ~~[\Sigma]_{\bk,\bp} = \delta_{\bk,\bp} \begin{pmatrix}
        0 & \sum_i\Gamma^i \Delta_i(\bk)\\
        \sum_i\Gamma^i \Delta_i^\dagger(\bk) & 0
    \end{pmatrix}
\end{equation}
The particle and hole Green's function are 
\begin{equation}
    \hat G_p(k)=[i\omega_n-\mathcal{H}(\bk)]^{-1}, ~~\hat G_h(k)=[i\omega_n+\mathcal{H}^T(-\bk)]^{-1}.
\end{equation}
By taking $\delta S[\Delta]/\delta \Delta_i^\dagger(\bp)=0$, and using
\begin{equation}
    (\mathcal{G}_0^{-1}+\Sigma)^{-1}\approx (1-\mathcal{G}_0\Sigma)\mathcal{G}_0
\end{equation}
we obtain the linearized gap equation
\begin{equation}
    \Delta_i(\bp)=-\int\frac{d^2\bk}{(2\pi)^2}\sum_{j}\tilde\chi_{\bp,\bk}P^{i,j}_{\bk}\Delta_j(\bk).\label{eq:lineargaps}
\end{equation}
where
 \begin{equation}
     P^{i,j}_{\bk}:=T\sum_n\text{Tr}[\hat G_p(k)\Gamma^j\hat G_h(k)\Gamma^i]=T\sum_n\text{Tr}[\hat G_h(k)\Gamma^i\hat G_p(k)\Gamma^j]
 \end{equation}
is even in $\bk$ since the model we considered has inversion symmetry. 
Eq.~\eqref{eq:lineargaps} is an eigenvalue problem and the leading pairing channel  will be the one whose eigenvalue first reaches $1$.

\subsection{2. Symmetry assisted simplification of the linearized gap equation} 
\label{sub:2_symmetry_assisted_simplification_of_the_linearized_gap_equation}

The gap equation can be further simplified by exploring the symmetry properties of the gap function $\Delta_i(\bk)$  under inversion, $\bk\to-\bk$, and under a shift by a primitive reciprocal lattice vector,  $\bk\to \bk+\bm{G}_{1,2}$, where 
\begin{equation}
    \bm{G}_1 = (2\pi,0), ~~\bm{G}_2=(0,2\pi).\label{eq:G12}
\end{equation}
The first property is manifest from Fermi statistics. For a generic Cooper pair in which two electrons have spin, orbital (or sublattice) and momentum indices, and if it is in even frequency pairing, the product of spin index interchange, orbital index interchange and $\bk\to-\bk$ must be $-1$. Since the properties of spin index interchange and orbital index interchange can be directly read from the $\Gamma^i$ matrices (e.g. spin-singlet pairing is odd under spin index interchange), it is straightforward to conclude whether a particular $\Delta_i(\bk)$ is even or odd under $\bk\to-\bk$. As for the symmetry properties under $\bk\to\bk+\bm{G}_{1,2}$, we note that intra-unit-cell pairings (those with $\tau_1$ or $\tau_2$) acquire an additional form factor, according to the property $\mathcal{H}_0(\bk + \bm{G}_{1,2}) = \tau_3 \mathcal{H}_0(\bk) \tau_3$,  which is odd under $\bk\to\bk+\bm{G}_{1,2}$, while others are even under this operation. These considerations lead to the results summarized in Table \ref{tab:SYM_2} for the eight attractive pairing channels. Below we employ these symmetry operations to further simplify the linearized gap equation.
\begin{table}
  \centering
  \begin{tabular}{lcccccccc}
    \hline\hline
       \text{operation}& ~$\Delta_1(\bk)$~ & ~$\Delta_2(\bk)$~ & ~$\Delta_3(\bk)$ ~&~ $\Delta_4(\bk)$ ~&~ $\Delta_5(\bk)$ ~&~ $\Delta_6(\bk)$ ~&~ $\Delta_7(\bk)$ ~&~ $\Delta_8(\bk)$\\\hline
       $\bk\to-\bk$ & $- $& $- $&$ -$ & $-$ & $+$ & $+$ & $-$ & $-$  \\
       $\bk\to\bk+\bm{G}_{1,2}$ & $+ $& $+ $&$+$ & $+$ & $-$ & $-$ & $-$ & $-$  \\
       \hline
  \end{tabular}
     \caption{Transformation properties of the gap functions  $\Delta_i(\bk)$ under two different symmetry operations: inversion and shift by a primitive reciprocal lattice vector. Here, $\bm{G}_{1,2}$ is defined in Eq.~\eqref{eq:G12} and $\pm$ is used to indicate that under the corresponding symmetry operation, the gap $\Delta_i(\bk)$ transforms into $\pm\Delta_i(\bk)$.}\label{tab:SYM_2}
  \end{table}


We first discuss the parity operation $\bk\to-\bk$, under which $\Delta_i(\bk)$ is either even or odd.
Generically, we can write
\begin{equation}
    \Delta_i(\bk)=\frac{\Delta_i(\bk)\pm\Delta_i(\bk)}{2}\label{eq:parityk}
\end{equation}
where $+$ is for $\Delta_i$ even in $\bk$ while $-$ is for $\Delta_i$ odd in $\bk$. 
One can show that $\Delta_i(\bk)$ with different parities do not mix. Indeed, using Eq.~\eqref{eq:parityk}, we manipulate Eq.~\eqref{eq:lineargaps} as 
\begin{equation}
    \begin{aligned}
        \Delta_i(\bp)&=-\int\frac{d^2\bk}{(2\pi)^2}\sum_{j}\tilde\chi_{\bp,\bk}P^{i,j}_{\bk}\frac{\Delta_j(\bk)\pm\Delta_j(-\bk)}{2},\\
        \Delta_i(-\bp)&=-\int\frac{d^2\bk}{(2\pi)^2}\sum_{j}\tilde\chi_{-\bp,\bk}P^{i,j}_{\bk}\frac{\Delta_j(\bk)\pm\Delta_j(-\bk)}{2},\\
        \frac{\Delta_i(\bp)\pm\Delta_i(-\bp)}{2}&=-\frac{1}{2}\int\frac{d^2\bk}{(2\pi)^2}\left(\sum_{j}\tilde\chi_{\bp,\bk}P^{i,j}_{\bk}\frac{\Delta_j(\bk)\pm\Delta_j(-\bk)}{2}\pm\sum_{j}\tilde\chi_{-\bp,\bk}P^{i,j}_{\bk}\frac{\Delta_j(\bk)\pm\Delta_j(-\bk)}{2}\right).
    \end{aligned}\label{eq:lineargap2}
\end{equation}
The last line is simply
\begin{equation}
    \frac{\Delta_i(\bp)\pm\Delta_i(-\bp)}{2}=-\frac{1}{4}\int\frac{d^2\bk}{(2\pi)^2}\sum_{j}\left[\left(\tilde\chi_{\bp,\bk}\pm\tilde\chi_{\bp,-\bk}\right)\pm\left(\tilde\chi_{-\bp,\bk}\pm\tilde\chi_{-\bp,-\bk}\right)\right]P^{i,j}_{\bk}\Delta_j(\bk).
\end{equation}
It can be inferred from the above equation that since $\tilde\chi_{\bp,\bk}=\tilde\chi_{-\bp,-\bk}$ and $\tilde\chi_{\bp,-\bk}=\tilde\chi_{-\bp,\bk}$, the even- and odd-$\bk$ pairings do not mix. 
As an example, if $\Delta_i$ is even in $\bk$ and $\Delta_j$ is odd in $\bk$, the kernel becomes $\left(\tilde\chi_{\bp,\bk}-\tilde\chi_{\bp,-\bk}\right)+\left(\tilde\chi_{-\bp,\bk}-\tilde\chi_{-\bp,-\bk}\right)=0$. And similarly if $\Delta_i$ is odd in $\bk$ and $\Delta_j$ is even in $\bk$, the kernel $\left(\tilde\chi_{\bp,\bk}+\tilde\chi_{\bp,-\bk}\right)-\left(\tilde\chi_{-\bp,\bk}+\tilde\chi_{-\bp,-\bk}\right)$ also vanishes. 
Therefore, the kernel survives only when $\Delta_i$ and $\Delta_j$ both have the same parity under $\bk\to-\bk$.

This implies we can consider the two subspaces separately. For $i,j\in$ even-in $\bk$ pairing, we have 
\begin{equation}
    \Delta_i(\bp)=-\int\frac{d^2\bk}{(2\pi)^2}\sum_{j}\frac{\tilde\chi_{\bp,\bk}+\tilde\chi_{\bp,-\bk}}{2}P^{i,j}_{\bk}\Delta_j(\bk).\label{eq:gap1}
\end{equation}
while for for $i,j\in$ odd-in $\bk$ pairing, we have
\begin{equation}
    \Delta_i(\bp)=-\int\frac{d^2\bk}{(2\pi)^2}\sum_{j}\frac{\tilde\chi_{\bp,\bk}-\tilde\chi_{\bp,-\bk}}{2}P^{i,j}_{\bk}\Delta_j(\bk).\label{eq:gap2}
\end{equation}

We can use a similar trick to exploit the symmetry of the gap functions with respect to the shift by $\bm{G}_{1,2}$.
 For later convenience, we note that $\tilde\chi_{\bp,\bk}$ depends on $J$ and the effect of shifting $\bk$ (or $\bp$) by either $\bm{G}_1$ or $\bm{G}_2$ is simply flipping the sign of $J$, see Eq. (\ref{eq:chirJ}). 
 Similarly, $P_{\bk}^{i,j}$ depends on $t$ (the NN hopping) and $\lambda$  (the SOC) through the Hamiltonian $\mathcal{H}_0$, and the effect of shifting $\bk$ in $\mathcal{H}_0$ by either $\bm{G}_1$ or $\bm{G}_2$ is simply flipping the signs of both $t$ and $\lambda$.
Here it is also straightforward to see that the gap functions that transform differently under $\bk\to\bk+\bm{G}_{1,2}$ do not mix, i.e., the kernel will vanish if $\Delta_i$ and $\Delta_j$ do not have the same property under such symmetry operation. 

Specifically, consider first  $i=5,6,7,8$, for which $\Delta_i(\bk)$ satisfies
\begin{equation}
    \Delta_i(\bk+\bm{G}_1)=\Delta_i(\bk+\bm{G}_2)=-\Delta_i(\bk)=-\Delta_i(\bk+\bm{G}_1+\bm{G}_2),
\end{equation}
Therefore, according to Table \ref{tab:SYM_2}, for $i,j\in{5,6,7,8}$ the gap equation is equivalent to
\begin{equation}
    \Delta_{i}(\bp)=-\frac{1}{2}\sum_{j,\bk}[V_{\bp,\bk}^{i,j}(J)P^{i,j}_{\bk}(t,\lambda)-V_{\bp,\bk+\bm{G}}^{i,j}(J)P^{i,j}_{\bk+\bm{G}}(t,\lambda)]\Delta_j(\bk),
\end{equation}
where $\bm{G}=\bm{G}_1$ or $\bm{G}_2$, and we will explicitly use the fact that
\begin{equation}
    V_{\bp,\bk+\bm{G}_{1,2}}^{i,j}(J)=V_{\bp,\bk}^{i,j}(-J),~~ P^{i,j}_{\bk+\bm{G}_{1,2}}(t,\lambda)=P^{i,j}_{\bk}(-t,-\lambda).
\end{equation}
The interaction $V_{\bp,\bk}^{i,j}(J)$ is given in Eq.~\eqref{eq:gap1} or \eqref{eq:gap2}, depending on whether $\Delta_{i,j}(\bk)$ is even or odd under $\bk\to-\bk$. Specifically,
\begin{equation}
    V_{\bp,\bk}^{i,j}(J)=\frac{\tilde\chi_{\bp,\bk}(J)+\tilde\chi_{\bp,-\bk}(J)}{2}
\end{equation}
if $i,j$ correspond to even-in-$\bk$ pairing ($i,j=1,2$); while
\begin{equation}
    V_{\bp,\bk}^{i,j}(J)=\frac{\tilde\chi_{\bp,\bk}(J)-\tilde\chi_{\bp,-\bk}(J)}{2}\label{eq:oddkV}
\end{equation}
if $i,j$ corresponds to odd-in-$\bk$ pairing ($i,j=3,4$).

On the other hand, for $i=1,2,3,4$, $\Delta_i(\bk)$ satisfies
\begin{equation}
    \Delta_i(\bk+\bm{G}_1)=\Delta_i(\bk+\bm{G}_2)=\Delta_i(\bk)=\Delta_i(\bk+\bm{G}_1+\bm{G}_2).
\end{equation}
Therefore, the gap equation for $i,j\in\{1,2,3,4\}$ is equivalent to
\begin{equation}
    \Delta_{i}(\bp)=-\frac{1}{2}\sum_{j,\bk}[V_{\bp,\bk}^{i,j}(J)P^{i,j}_{\bk}(t,\lambda)+V_{\bp,\bk+\bm{G}}^{i,j}(J)P^{i,j}_{\bk+\bm{G}}(t,\lambda)]\Delta_j(\bk)
\end{equation}
where $V_{\bp,\bk}^{i,j}(J)$ is given in Eq.~\eqref{eq:oddkV} since the pairing here should be all odd-in $\bk$ according to Table \ref{tab:SYM_2}. The above analysis is nothing but to write the kernel matrix of the linearized gap equation in a block-diagonalized way based on the their known symmetries. 

To summarize, if we write the linearized gap equation as
\begin{equation}
    \Delta_i(\bp)=-\sum_{\bk,j}K_{\bp,\bk}^{i,j}\Delta_j(\bp)\label{eq:gapeqsymm1}
\end{equation}
 and define three symmetrized interactions
\begin{equation}
    \begin{aligned}
        \begin{aligned}
            \tilde V^a_{i,j;\bp,\bk}= \frac{[\tilde\chi_{\bp,\bk}(J)-\tilde\chi_{\bp,-\bk}(J)]P^{i,j}_{\bk}(t,\lambda)+[\tilde\chi_{\bp,\bk}(-J)-\tilde\chi_{\bp,-\bk}(-J)]P^{i,j}_{\bk}(-t,-\lambda)}{4},\\
            \tilde V^b_{i,j;\bp,\bk}=\frac{[\tilde\chi_{\bp,\bk}(J)+\tilde\chi_{\bp,-\bk}(J)]P^{i,j}_{\bk}(t,\lambda)-[\tilde\chi_{\bp,\bk}(-J)+\tilde\chi_{\bp,-\bk}(-J)]P^{i,j}_{\bk}(-t,-\lambda)}{4}\\
            \tilde V^c_{i,j;\bp,\bk}= \frac{[\tilde\chi_{\bp,\bk}(J)-\tilde\chi_{\bp,-\bk}(J)]P^{i,j}_{\bk}(t,\lambda)-[\tilde\chi_{\bp,\bk}(-J)-\tilde\chi_{\bp,-\bk}(-J)]P^{i,j}_{\bk}(-t,-\lambda)}{4}\\   
        \end{aligned}
    \end{aligned}\label{eq:gapeqsymm2}
\end{equation}
then the kernel matrix in $i,j$ space is given by
\begin{equation}
   K_{\bp,\bk}^{i,j}=\begin{pmatrix}
        \begin{pmatrix}
        \tilde V^a_{1,1;\bp,\bk} &  \tilde V^a_{1,2;\bp,\bk} &  \tilde V^a_{1,3;\bp,\bk} &  \tilde V^a_{1,4;\bp,\bk}\\
        \tilde V^a_{2,1;\bp,\bk} &  \tilde V^a_{2,2;\bp,\bk} &  \tilde V^a_{2,3;\bp,\bk} &  \tilde V^a_{2,4;\bp,\bk}\\
        \tilde V^a_{3,1;\bp,\bk} &  \tilde V^a_{3,2;\bp,\bk} &  \tilde V^a_{3,3;\bp,\bk} &  \tilde V^a_{3,4;\bp,\bk}\\
        \tilde V^a_{4,1;\bp,\bk} &  \tilde V^a_{4,2;\bp,\bk} &  \tilde V^a_{4,3;\bp,\bk} &  \tilde V^a_{4,4;\bp,\bk}\\
    \end{pmatrix} & 0 & 0 \\
     0 & \begin{pmatrix}
        \tilde V^b_{5,5;\bp,\bk} & \tilde V^b_{5,6;\bp,\bk}\\
        \tilde V^b_{6,5;\bp,\bk} & \tilde V^b_{6,6;\bp,\bk}
    \end{pmatrix}  & 0\\
    0 & 0 & \begin{pmatrix}
        \tilde V^c_{7,7;\bp,\bk} & \tilde V^c_{7,8;\bp,\bk}\\
        \tilde V^c_{8,7;\bp,\bk} & \tilde V^c_{8,8;\bp,\bk}
    \end{pmatrix} \\
   \end{pmatrix}.\label{eq:gapeqsymm3}
\end{equation}
We next solve the linearized gap equation in Eq.\eqref{eq:gapeqsymm1}, which is equivalent to diagonalizing the matrix $-K_{\bp,\bk}^{i,j}$. An eigenvalue corresponds to the effective pairing strength in a certain channel, and the corresponding eigenvector is the pairing gap function. The leading pairing channel is the one which has the largest eigenvalue, and the temperature at which the largest eigenvalue reaches 1 gives the transition temperature $T_c$. Our focus will be to determine the leading pairing channel, instead of determining $T_c$. 

\section{III. Eigenvectors of the gap function in momentum space} 
\label{sec:eigenvectors_of_the_gap_function_in_momentum_space}

We proceed to show representative examples of the momentum space eigenvectors obtained from solving the linearized gap equation Eq.~\eqref{eq:gapeqsymm1}, obtaining the results discussed in the main text. Here we will always choose $t=1, t_a'=-0.3, t_b'=0.2$ and no SOC. Different leading pairings are achieved by tuning the chemical potential $\mu$ and $r/J$.

\subsection{1. $s'$-, $d'$-, and $p'$-wave} 
\label{sub:_sdp}

\begin{figure}
  \includegraphics[width=18cm]{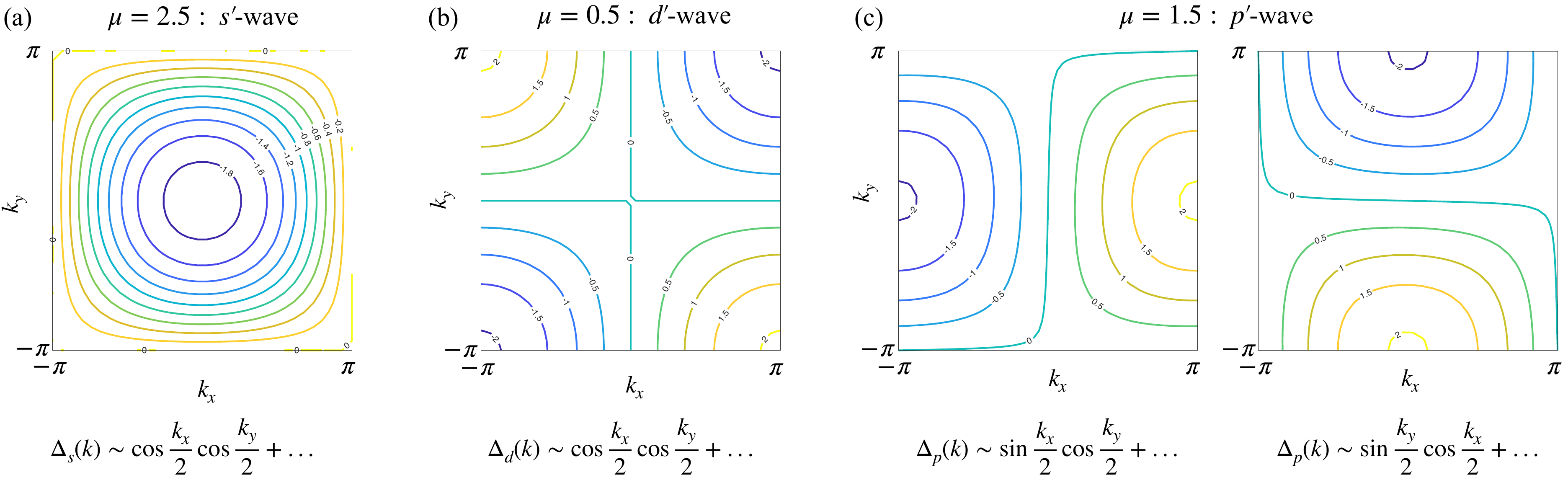}
  \caption{Gap functions for $s'$-, $d'$- and $p'$-wave pairing channels obtained numerically by diagonalizing the gap equation. The labels indicate the leading lattice harmonic for each case.}
  \label{fig:sdpeigenfunction}
\end{figure}

 We choose a representative $r/J=1.5$. The eigenfunctions of the leading pairing channels from diagonalizing the kernel in Eq.~\eqref{eq:gapeqsymm3} at $\mu=0.5, 1.5$ and $2.5$ are presented in Figs. \ref{fig:sdpeigenfunction}. We find the following results:
\begin{enumerate}
  \item For $\mu=0.5$, the leading channel eigenfunction has finite contribution from only $\Gamma^5$, and the corresponding $\Delta_1(\bk)$ is shown in Fig. \ref{fig:sdpeigenfunction}(b). It almost coincides with the $d'$-wave form factor $\Delta_d(\bk)\sim\sin\frac{k_x}{2}\sin\frac{k_y}{2}$. 
  \item For $\mu=2.5$, the leading channel eigenfunction has finite contribution from only $\Gamma^5$ as well, with the corresponding $\Delta_1(\bk)$ shown in Fig. \ref{fig:sdpeigenfunction}(a). It almost coincides with the $s'$-wave form factor $\Delta_s(\bk)\sim\cos\frac{k_x}{2}\cos\frac{k_y}{2}$.
  \item For $\mu=1.5$, the leading channel has a twofold degeneracy, and the two degenerate eigenfunctions both have finite contribution from only the $\Gamma^7$ channel. The corresponding $\Delta_3(\bk)$ are shown in Fig. \ref{fig:sdpeigenfunction}(c). The two eigenfunctions almost coincide with the $p'_x$-wave form factor $\Delta_{p_x}(\bk)\sim \sin\frac{k_x}{2}\cos\frac{k_y}{2}$ and the $p'_y$-wave form factor $\Delta_{p_y}(\bk)\sim \sin\frac{k_y}{2}\cos\frac{k_x}{2}$. Their symmetric and anti-symmetric combinations give the $p'$-wave form factors  $u_+(\bk)=\sin[(k_x + k_y)/2] $ and $u_-(\bk)=\sin[(k_x - k_y)/2]$ introduced in the main text.
\end{enumerate}
For other parts in the phase diagram of Fig. 1 of the main text, the profile of the pairing eigenfunctions remain almost the same.

\begin{figure}
  \includegraphics[width=0.9\linewidth]{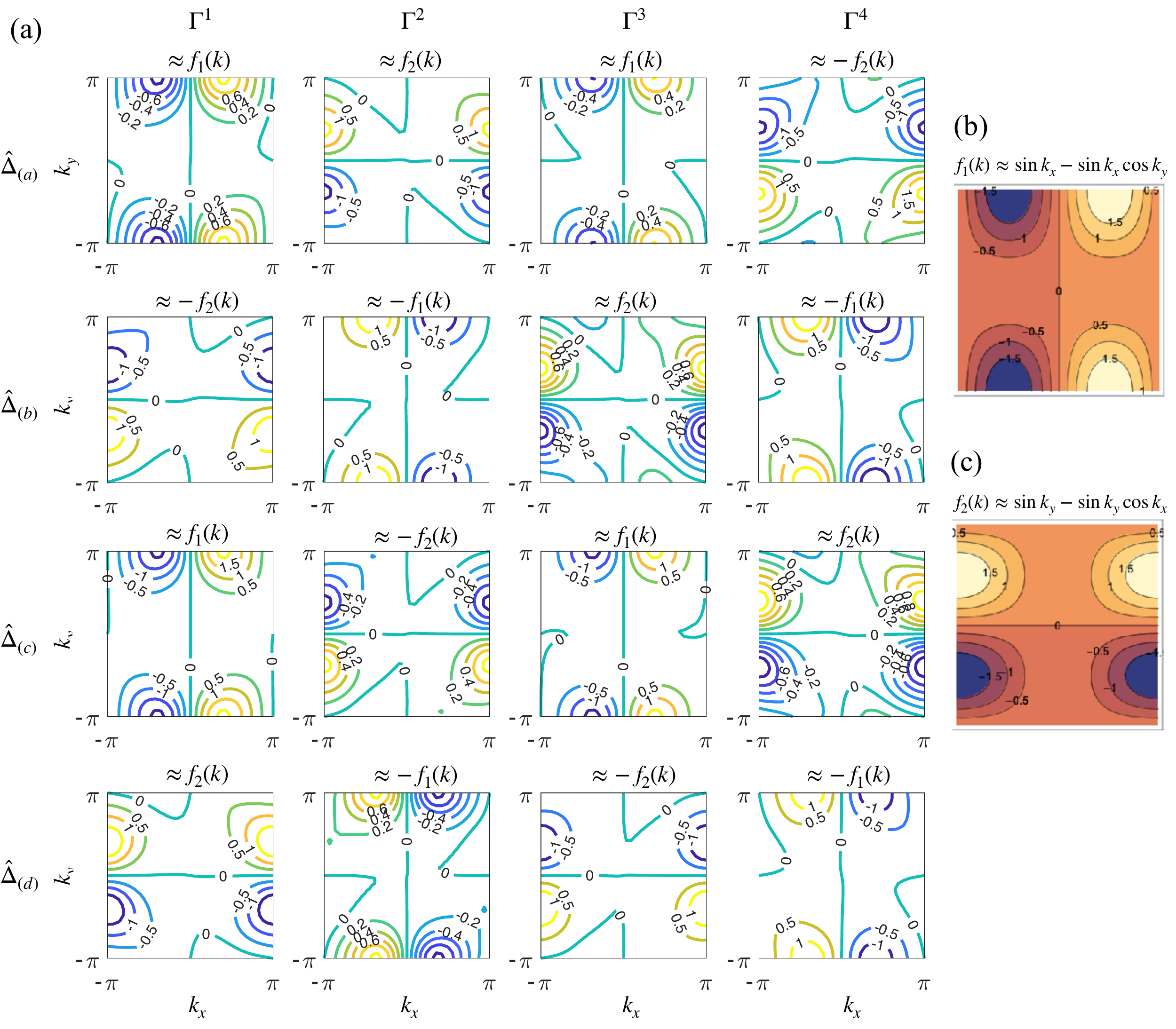}
  \caption{(a) Gap functions for the $p$-wave pairing channel, which has four degenerate eigenfunctions denoted by $\hat\Delta_{(a)},\hat\Delta_{(b)},\hat\Delta_{(c)},\hat\Delta_{(d)}$. Each pairing state has nonzero contributions from four different $\Gamma^i$ with $i=1,2,3,4$, and each of them is dominated by either $f_1(\bk)$ or $f_2(\bk)$, which are shown in (b) and (c).}
  \label{fig:ppeigenfunction}
\end{figure}

\subsection{2. $p$-wave} 
\label{sub:_p}

When $r/J$ gets closer to $1$, another $p$-wave SC  arises, which has nonzero contributions from all $\Gamma^{1-4}$ channels. This state has fourfold degeneracy as noted in the main text. 
In Fig. \ref{fig:ppeigenfunction}(a) we present one example of these eigenfunctions obtained from diagonalizing Eq.\eqref{eq:gapeqsymm3} with $r/J=1.0753$ and $\mu=0.75$. By comparing these functions with two lattice harmonics 
\begin{equation}
  \begin{aligned}
    f_1(\bk)=\sin k_x - \sin k_x \cos k_y,\\
    f_2(\bk)=\sin k_y - \sin k_y \cos k_x,
  \end{aligned}\label{eq:twoapproxharmonics}
\end{equation}
which is plotted in Fig. \ref{fig:ppeigenfunction}(b) and (c),
we see that all the eigenfunctions in Fig. \ref{fig:ppeigenfunction}(a) have dominant contributions from either $f_1(\bk)$ or $f_2(\bk)$. If we approximately consider only the two harmonics in Eq. \eqref{eq:twoapproxharmonics}, we are able to identify in Fig. \ref{fig:ppeigenfunction}(a) which eigenfunction corresponds to $f_1(\bk)$ and which corresponds to $f_2(\bk)$. This information is presented on the top of each plot. Denoting the four degenerate pairing structure from diagonalizing Eq. \eqref{eq:gapeqsymm3} by $\hat\Delta_{(a)}$, $\hat\Delta_{(b)}$, $\hat\Delta_{(c)}$ and $\hat\Delta_{(d)}$, we then have
\begin{equation}
  \begin{aligned}
    &\hat\Delta_{(a)}\approx f_1(\bk)(\Gamma^1+\Gamma^3)+f_2(\bk)(\Gamma^2-\Gamma^4)\\
    &\hat\Delta_{(b)}\approx -f_1(\bk)(\Gamma^2+\Gamma^4)-f_2(\bk)(\Gamma^1-\Gamma^3)\\
    &\hat\Delta_{(c)}\approx f_1(\bk)(\Gamma^1+\Gamma^3)-f_2(\bk)(\Gamma^2-\Gamma^4)\\
    &\hat\Delta_{(d)}\approx -f_1(\bk)(\Gamma^2+\Gamma^4)+f_2(\bk)(\Gamma^1-\Gamma^3)\\
  \end{aligned}\label{eq:abcd}
\end{equation}
Since these four states are degenerate, we can take linear combinations of these states and the result is still an eigenstate. Thus, for later convenience we introduce a new set of four states indexed by $j$, which are obtained from proper combinations of Eq. \eqref{eq:abcd}, namely
\begin{equation}
    \begin{aligned}
       \hat\Delta_{(c)}-\hat\Delta_{(b)}\to~~ &\hat\Delta_{(1)}(\bk)=\Delta_{(1)}[(\Gamma^1+\Gamma^4)v_+(\bk)+(\Gamma^2+\Gamma^3)v_-(\bk)],\\
      \hat\Delta_{(a)}-\hat\Delta_{(d)}\to~~ &\hat \Delta_{(2)}(\bk)=\Delta_{(2)}[(\Gamma^1+\Gamma^4)v_-(\bk)+(\Gamma^2+\Gamma^3)v_+(\bk)],\\
        \hat\Delta_{(a)}+\hat\Delta_{(d)}\to~~  &\hat\Delta_{(3)}(\bk)=\Delta_{(3)}[(\Gamma^1-\Gamma^4)v_+(\bk)-(\Gamma^2-\Gamma^3)v_-(\bk)],\\
         \hat\Delta_{(b)}+\hat\Delta_{(c)}\to~~  &\hat\Delta_{(4)}(\bk)=\Delta_{(4)}[(\Gamma^1-\Gamma^4)v_-(\bk)-(\Gamma^2-\Gamma^3)v_+(\bk)].\\
    \end{aligned}\label{eq:fourcomponents2}
\end{equation}
Here, 
\begin{equation}
  v_\pm(\bk)=f_1(\bk)\pm f_2(\bk)=(\sin k_x\pm\sin k_y)-\sin(k_x \pm k_y).
\end{equation}
Using the definitions of the $\Gamma^i$ matrices, one directly arrives at Eq. (8) in the main text. Physically, $\hat\Delta_{(1)}$ and $\hat\Delta_{(2)}$ correspond to spin-up pairings, while $\hat\Delta_{(3)}$ and $\hat\Delta_{(4)}$ correspond to spin-down pairings. Note that although here we have neglected higher order harmonics, which come from longer range Cooper pair separation on a lattice, these contributions may become important when the system is very close to the critical point $r/J\to1$. What we have shown here is that, as long as the system is not very close to the critical point, the pairing is still dominated by  a few leading harmonics. 

\subsection{3. Impact of finite SOC} 
\label{sub:3_with_finite_soc}

\begin{figure}
  \includegraphics[width=0.9\linewidth]{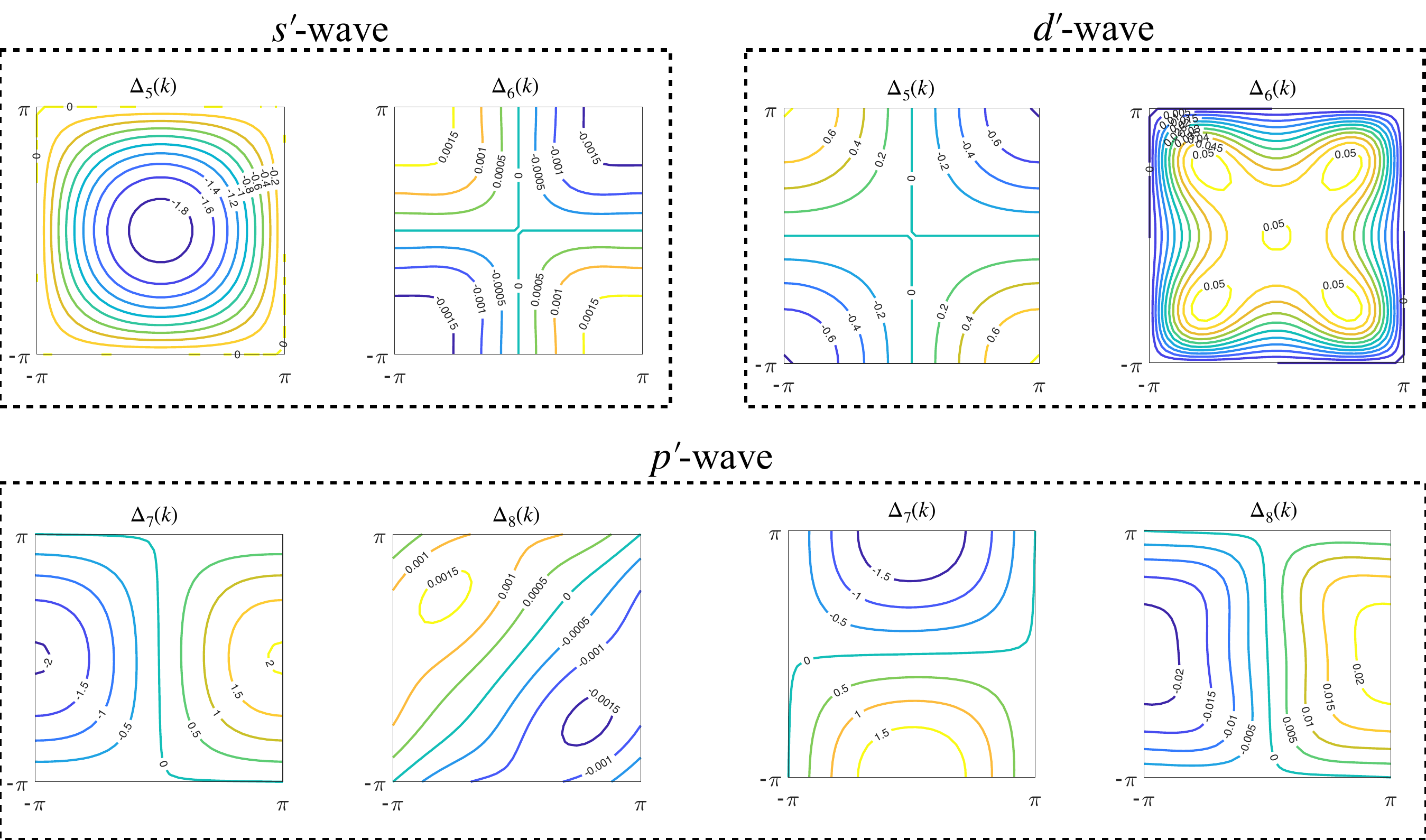}
  \caption{Eigenfunctions of the $s'$-, $d'$- and $p'$-wave pairing in the presence of a finite SOC ($\lambda=0.3$). Other parameters are identical to those used in Fig. \ref{fig:sdpeigenfunction}. In all cases we observe $\Delta_6(\bk)\ll\Delta_5(\bk)$ and $\Delta_8(\bk)\ll\Delta_7(\bk)$, so that even in the presence of SOC the pairings are still dominated by $\Delta_5(\bk)$ and $\Delta_7(\bk)$.}\label{fig:withSOC}
\end{figure}

We now discuss how a finite SOC $\lambda$ impacts the eigenvectors. As explained in the main tetx, we consider a Kane-Mele type SOC which takes the following form
\begin{equation}
  h_2(\bk)\tau_2\sigma_3=\lambda\sin\frac{k_x}{2}\sin\frac{k_y}{2}\tau_2\sigma_3.
\end{equation}
This term preserves time-reversal and four-fold rotation symmetry. It opens up a gap in the non-interacting band structure at $M=(\pi,\pi)$.
We will see the effect of $\lambda\neq0$ is to introduce mixtures between $\Gamma^5$ and $\Gamma^6$, and also between $\Gamma^7$ and $\Gamma^8$, as allowed by the symmetry of the gap equations \eqref{eq:gapeqsymm3}. In Fig. \ref{fig:withSOC} we present the eigenfunctions for $s'$-, $d'$- and $p'$-wave pairings in the presence of $\lambda=0.3$. For convenience of comparison, we keep all other parameters the same as used in generating Fig. \ref{fig:sdpeigenfunction}.

It is clear from the numerical results that for both the $s'$-wave and $d'$-wave cases, a finite SOC also induces a nonzero contribution from the $\Gamma^6$ channel. However, in both cases we find $\Delta_5(\bk)\gg\Delta_6(\bk)$, namely, the Cooper pair configuration is predominantly from the $\Gamma^5$ channel. Note that, in the $s'$-wave case, the new induced gap function has the form $\hat{\Delta}_6(\bk)=\tilde{\Delta}_s \sin\frac{k_x}{2}\sin\frac{k_y}{2} \tau_2 \sigma_1$, corresponding to an even-parity spin-triplet state. Despite the $d$-wave character of the form factor, the gap function is $s$-wave due to the transformation properties of the spin and sublattice Pauli matrices under the point group symmetries, as we explain in the next section (see Table \ref{table:SYM3}). 

Similar results happen to the $p'$-wave pairing in SOC is included. Without SOC, as we showed above, the two degenerate pairings have nonzero contribution only from the $\Gamma^7$ channel. In the presence of SOC, both degenerate states have finite contributions from the $\Gamma^8$ channel as well. But $\Delta_8(\bk)\ll\Delta_7(\bk)$, so the pairing is predominantly $\Delta_7(\bk)$ even with SOC.

For the $p$-wave pairing, a finite SOC does not introduce nonzero components in a new channel other than $\Gamma^{1,2,3,4}$. However, this SOC will split the four-fold degeneracy into $2\times$ two-fold degeneracy. 

\subsection{4. Symmetries of the intra-unit cell pairing} 
\label{sub:3_symmetries_of_the_intra_unit_cell_pairing}

We finish this section by explicitly stating the symmetries of the intra-unit-cell gap functions corresponding to the attractive pairing channels. The Hamiltonian in Eq. (1) of the main text has several symmetries, including time-reversal $\mathcal{T}$ and fourfold rotation $C_4$,  which we list in Table \ref{table:SYM3}.  The transformation of the AM and the SC order parameters under these transformations can be straightforwardly obtained, which we present in the same Table. One important subtlety is that due to the particle-particle nature of the pairing order, symmetry operations ($S$) act on them in the spin/sublattice subspace as 
$\tau_i \sigma_j \to S \tau_i \sigma_j S^T$, rather than $\tau_i \sigma_j \to S \tau_i \sigma_j S^\dag$. 

In particular, we see that the AM order parameter transforms as the $B_{2g}$ irreducible representation, the $s'$-wave pairing order parameter transforms as $A_{1g}$, and the $d'$-wave pairing order parameter, as $B_{2g}$. As mentioned in the previous section, we also confirm that the $\tilde{s}'$-wave state induced by SOC also transforms as the $A_{1g}$ irreducible representation. 

\begin{table}[h]
  \centering
  \begin{tabular}{cccccccc}
    \hline\hline
       &$\mathcal{T}$ & $C_4$ & $\mathcal{I}$ & $\mathcal{M}_{x}$ & $\mathcal{M}_{y}$ & $\mathcal{M}_{x+y}$ & $\mathcal{M}_{x-y}$ \\\hline
       &$i\sigma_2K$ & $\tau_1$ & id & $i\sigma_1$ & $i\sigma_2$ & $i\frac{\sigma_1+\sigma_2}{\sqrt{2}}\tau_1$& $i\frac{\sigma_1-\sigma_2}{\sqrt{2}}\tau_1$ \\
       \hline 
       $N\sim \tau_3\sigma_3$ &$-$&$-$&$+$&$-$&$-$&$+$&$+$  \\
       $\Delta_d \sim \sin(k_x/2) \sin(k_y/2)\tau_1 i\sigma_2$ &$+$&$-$&$+$&$-$&$-$&$+$&$+$  \\
       $\Delta_s \sim \cos(k_x/2) \cos(k_y/2)\tau_1 i\sigma_2$&$+$&$+$&$+$&$+$&$+$&$+$&$+$\\
       $\tilde\Delta_s \sim \sin(k_x/2) \sin(k_y/2)\tau_2 \sigma_1$&$+$&$+$&$+$&$+$&$+$&$+$&$+$ \\
       \hline
  \end{tabular}
     \caption{In the second row, we list the representation of the symmetries of the Lieb lattice model in the combined subspace of spin and sublattice subspace. Here, $\mathcal{M}_i$ denotes a reflection with respect to a mirror plane that is perpendicular to the $i$-axis. In the next three rows, we present the symmetry transformations of the order parameters discussed in the main text (we assume the pairing order parameters are real). Note that this analysis also includes the transformation in $\bk$-space.}
     \label{table:SYM3}
  \end{table}

\section{IV. Ginzburg-Landau analysis for the degenerate SC states} 
\label{sec:gl_analysis_for_the_degenerate_sc_states}
In this section, we employ a Ginzburg-Landau (GL) free energy analysis of the degenerate $p'$-wave and $p$-wave  pairing channels, showing that they naturally lead to intrinsic topological superconductivity. 

\subsection{1. Chiral $p'$-wave SC} 
\label{sub:chiral_sc}

We first discuss the $p'$-wave state, which has nonzero contribution from only the $\Gamma^7$ channel. As discussed in the main text, the twofold degenerate gap function can be written as 
\begin{equation}
  \hat\Delta(\bk)=\Gamma^7[\Delta_{(1)}u_+(\bk)+\Delta_{(2)}u_-(\bk)],
\end{equation}
where $u_+(\bk)\sim \sin[(k_x+k_y)/2]$ and $u_-(\bk)\sim \sin[(k_x-k_y)/2]$. The effective GL free energy is obtained by expanding the SC action in Eq. \eqref{eq:tracelog} in powers of $\hat\Delta$ in a certain channel. Here, for the purpose of identifying the state which leads to the lowest free energy, it suffices to keep the terms up to the quartic order and we obtain
\begin{equation}
  \mathcal{F}_{GL}=\alpha\left(|\Delta_{(1)}|^2+|\Delta_{(2)}|^2\right)+\beta_1\left(|\Delta_{(1)}|^4+|\Delta_{(2)}|^4\right)+\beta_2|\Delta_{(1)}|^2|\Delta_{(2)}|^2+\beta_3\left(\Delta_{(1)}^2(\Delta_{(2)}^*)^2+\Delta_{(2)}^2(\Delta_{(1)}^*)^2\right)+...
\end{equation}
Here $\alpha$ determines $T_c$ of this pairing state, and higher order terms are all omitted. 
The third term with coefficient $\beta_3$ determines whether the state favors spontaneous time-reversal symmetry breaking. We find:
\begin{equation}
  \beta_3=T\sum_{\omega_n}\int\frac{d^2\bk}{(2\pi)^2}\frac{4 u_+^2(\bk)u_-^2(\bk) \left(h_0^4+2 h_0^2 \left(3 h_1^2-h_3^2+\omega_n ^2\right)+h_1^4+h_1^2 \left(2 \omega_n ^2-6 h_3^2\right)+h_3^4-6 h_3^2 \omega_n ^2+\omega_n ^4\right)}{\left(-h_0^2-2 i h_0 \omega_n +h_1^2+h_3^2+\omega_n ^2\right)^2 \left(-h_0^2+2 i h_0 \omega_n +h_1^2+h_3^2+\omega_n ^2\right)^2}.
\end{equation}
Here, $\omega_n=(2n+1)\pi T$ is a fermionic Matsubara frequency, and $h_i=h_i(\bk)$ are the Hamiltonian components given in the main text. By explicit calculation we find $\beta_3>0$, implying a $\pi/2$ mod $\pi$ phase difference between the two components $\Delta_{(1)}$ and $\Delta_{(2)}$. This is the chiral $p'\pm ip'$ superconductivity discussed in the main text. Assuming that the two components have the same gap amplitude, the corresponding BdG Hamiltonian for this chiral SC state is 
\begin{equation}
    \begin{aligned}
        \mathcal{H}_\text{BdG}(\bk)&=\eta_3\tau_0\sigma_0 h_0(\bk)+\eta_3\tau_1\sigma_0 h_1(\bk)+\eta_3\tau_3\sigma_0 h_3(\bk)\\
        &+\Delta[\eta_1\tau_1\sigma_1u_+(\bk)+\eta_2\tau_1\sigma_1u_-(\bk)].\\
    \end{aligned}
\end{equation}
In this expression, we used $\eta_{1,2,3}$ to denote the Pauli matrices in the Nambu spinor space. The spectrum obtained from diagonalizing this Hamiltonian is fully gapped. 
 There are four such quasiparticle bands, and each of them is twofold degenerate. To show the topological character of this chiral SC state, we calculate the total Chern number for all the bands below Fermi level using the non-Abelian Berry connection\cite{FukuiHatsugaiSuzuki}. Indeed, the total Chern number found is $C=2$, indicating that two chiral modes can develop at the boundary. This pairing state thus belongs to class D, and is topologically non-trivial\cite{topoclassification}.

\subsection{2. Helical $p$-wave SC} 
\label{sub:helical_sc}

 Next, we discuss the $p$-wave state, which has nonzero contributions from all $\Gamma^{1-4}$ channels and is fourfold degenerate. A generic pairing state is the linear combination of the four degenerate states in Eq. \eqref{eq:fourcomponents2}, namely
\begin{equation}
    \begin{aligned}
       \hat\Delta(\bk)&=\Delta_{(1)}[(\Gamma^1+\Gamma^4)v_+(\bk)+(\Gamma^2+\Gamma^3)v_-(\bk)]\\
      &+\Delta_{(2)}[(\Gamma^1+\Gamma^4)v_-(\bk)+(\Gamma^2+\Gamma^3)v_+(\bk)]\\
      &+\Delta_{(3)}[(\Gamma^1-\Gamma^4)v_+(\bk)-(\Gamma^2-\Gamma^3)v_-(\bk)]\\
      &+\Delta_{(4)}[(\Gamma^1-\Gamma^4)v_-(\bk)-(\Gamma^2-\Gamma^3)v_+(\bk)]\\
    \end{aligned}
\end{equation}
where, as we discussed in the main text, $v_\pm(\bk)\approx (\sin k_x \pm\sin k_y)-\sin(k_x\pm k_y) $.  $\Delta_{(1)}$ and $\Delta_{(2)}$ correspond to spin-up pairing, while $\Delta_{(3)}$ and $\Delta_{(4)}$ correspond to spin-down pairing. 
 The GL free energy expanded up to quartic terms is obtained in a similar way as the $p'$-wave case, yielding:
\begin{equation}
  \begin{aligned}
    \mathcal{F}_{GL}=&\alpha\left(|\Delta_{(1)}|^2+|\Delta_{(2)}|^2+|\Delta_{(3)}|^2+|\Delta_{(4)}|^2\right)\\
    +&\beta_{\uparrow1}\left(|\Delta_{(1)}|^4+|\Delta_{(2)}|^4\right)+\beta_{\uparrow2}|\Delta_{(1)}|^2|\Delta_{(2)}|^2+\beta_{\uparrow3}\left(\Delta_{(1)}^2(\Delta_{(2)}^*)^2+\Delta_{(2)}^2(\Delta_{(1)}^*)^2\right)\\
    +&\beta_{\downarrow1}\left(|\Delta_{(3)}|^4+|\Delta_{(4)}|^4\right)+\beta_{\downarrow2}|\Delta_{(3)}|^2|\Delta_{(4)}|^2+\beta_{\downarrow3}\left(\Delta_{(3)}^2(\Delta_{(4)}^*)^2+\Delta_{(4)}^2(\Delta_{(3)}^*)^2\right)+...
  \end{aligned}
\end{equation}
From this expression, it is clear that spin-up and spin-down gaps are decoupled. Moreover, we confirmed from a direct calculation that 
\begin{equation}
  \beta_{\uparrow1}=\beta_{\downarrow1},~~\beta_{\uparrow2}=\beta_{\downarrow2},~~\beta_{\uparrow3}=\beta_{\downarrow3}
\end{equation}
and all of the coefficients are positive. This implies the phase differences between $\Delta_{(1)}$ and $\Delta_{(2)}$, and between $\Delta_{(3)}$ and $\Delta_{(4)}$ are both $\pi/2$ mod $\pi$. However, the four fold degeneracy gives rise to two possible ground states. In one, both the spin-up and spin-down sectors form a $p+ip$ state (same chirality), and in the other, the spin-up sector forms a $p+ip$ state while the spin-down sector forms a $p-ip$. In the first case, the time-reversal symmetry is spontaneously broken, and the system can be viewed as two copies of conventional chiral $p+ip$ superconductivity. In the second case, the time-reversal symmetry is preserved. Interestingly, this state is still topologically non-trivial in a sense that one can define a topological index $\mathbb{Z}_2=(C_\uparrow-C_\downarrow)/2=1$. This type of superconductor can host a pair of helical edge states. The corresponding BdG Hamiltonian is 
\begin{equation}
    \begin{aligned}
        \mathcal{H}_\text{BdG}(\bk)&=\eta_3\tau_0\sigma_0 h_0(\bk)+\eta_3\tau_1\sigma_0 h_1(\bk)+\eta_3\tau_3\sigma_0 h_3(\bk)\\
        &+\Delta[\eta_1\tau_0\sigma_0v_+(\bk)+\eta_2\tau_0\sigma_3v_-(\bk)\\
        & ~~~ +\eta_1\tau_3\sigma_0v_-(\bk)+\eta_2\tau_3\sigma_3v_+(\bk)].
    \end{aligned}\label{eq:helicalSC}
\end{equation}
which belongs to class DIII\cite{topoclassification}.



\section{V.  Phase diagrams} 
\label{sec:iii_eigenfunctions_and_phase_diagrams}

\begin{figure} [h]
    \includegraphics[width=16cm]{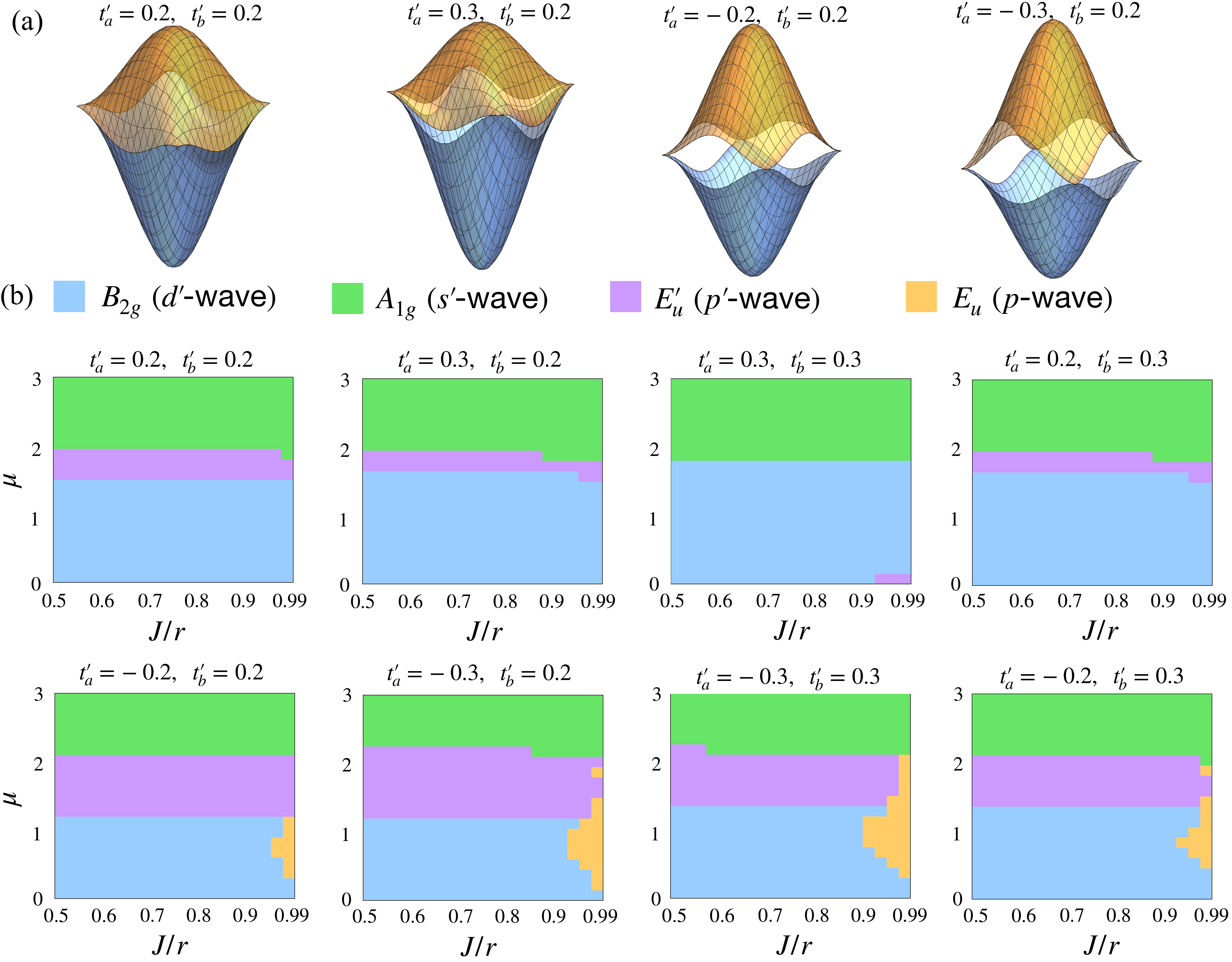}
    \caption{(a) Band structure for four different choices of $t'_a$ and $t'_b$ without SOC. The band gap closes at $M=(\pi,\pi)$ point. (b) Phase diagrams of the leading pairing instabilities obtained by numerically solving the linearized gap equations for various $t'_a$ and $t'_b$ values without SOC.}\label{fig:without_soc_}
\end{figure}

\begin{figure} [h]
    \includegraphics[width=16cm]{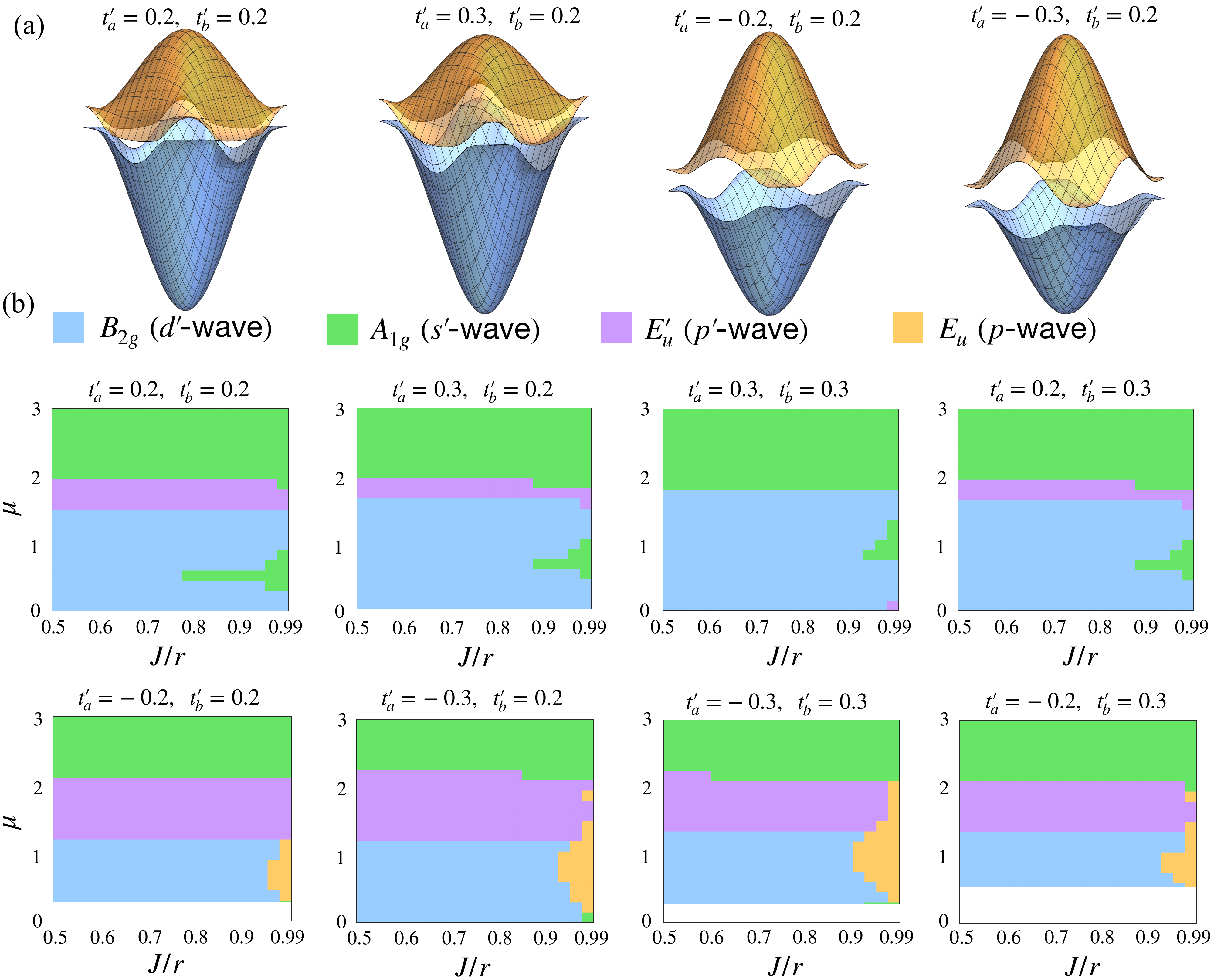}
    \caption{(a) Band structure for four different choices of $t'_a$ and $t'_b$ in the presence of a finite SOC ($\lambda=0.3$). Phase diagrams of the leading pairing instabilities obtained by numerically solving the linearized gap equations for various $t'_a$ and $t'_b$ values with SOC ($\lambda=0.2$).}\label{fig:with_soc_lambda_0_2_}
\end{figure}

\subsection{1. Without SOC} 
\label{sub:1_without_soc}

Fig. 1(c) of the main text shows the superconducting phase diagram obtained by solving Eqs. \eqref{eq:gapeqsymm1}, \eqref{eq:gapeqsymm2} and \eqref{eq:gapeqsymm3}, and investigating the eigenvector $\Delta_i(\bk)$ that corresponds to he largest eigenvalue. The set of parameters used in that figure were $t=1$, $t_a'=-0.3$, $t'_b=0.2$, and $\lambda=0$. Here, we obtain similar phase diagrams but for a different set of parameters.

Fig. \ref{fig:without_soc_} (a) presents band structures for various $t'_a$ and $t'_b$ values, and Fig. \ref{fig:without_soc_} (b) shows the results of the leading pairing as a function of $\mu$ and $J/r$. The phase diagrams can be roughly divided into two categories. When the two next-nearest-neighbor hopping parameters have similar magnitudes and the same sign (here we choose $t'_a,t'_b>0$), we do not find a $p$-wave pairing solution.  When $t'_a$ and $t'_b$ have similar magnitudes and opposite signs, the significant difference is the emergence of $p$-wave pairing near $\mu\sim1$ and $J/r>0.9$. 

To understand these results, we note that the term in the Hamiltonian of Eq. (1) of the main text that captures the full symmetry of the Lieb lattice and distinguishes it from a folded square lattice is the $h_3(\bk)$ term, with coefficient $(t_a' - t_b')$ \cite{antonenko2024mirrorchernbandsweyl}. Therefore, if $t_a' = t_b'$, the model actually describes a folded N\'eel antiferromagnet, whose fluctuations should favor singlet pairing. In this case, the $p$-wave state should be strongly suppressed, since the ground state is not similar to a spin-triplet Pomeranchuk state. This is consistent with the results of Fig. \ref{fig:without_soc_} (b), which show a strong suppression of the $p$-wave instability when  $|t_a'-t_b'|/t$ is small. 

\subsection{2. With SOC} 
\label{sub:inclusion_of_soc}


Fig. \ref{fig:with_soc_lambda_0_2_} presents the results with the SOC term added for comparison. Because there is a gap in the band structure at the $M$ point due to the finite SOC, no Fermi surface is present when the chemical potential is inside the gap, which requires a threshold chemical potential value to obtain a pairing state. The phase diagrams when $t'_a,t'_b>0$ are similar to those in Fig.\ref{fig:without_soc_}, despite the fact that their corners at small $\mu$ and large $J/r$ values are now dominated by $s'$-wave pairing. The phase diagrams when $t'_a<0$ and $t'_b>0$ are almost the same as those in Fig.\ref{fig:without_soc_}.



\section{VI. Fermi-surface projected $d'$-wave gap function}
\label{sec:vi}

In this section we show that when the Fermi surface (FS) is centered at the $M$ point, the $d'$-wave pairing state leads to point nodes at the Brillouin zone (BZ) boundary. For a small FS, we expand the tight-binding Hamiltonian near the $M$ point (setting $\lambda = 0$)
\begin{align}
\mathcal H_M(\bk)=&\,
\Bigl[-\delta\mu-\frac{t_a'+t_b'}{2}\bigl(k_x^2+k_y^2\bigr)\Bigr]\tau_0\sigma_0
-\,t\,k_xk_y\,\tau_1\sigma_0
-\,\frac{t_a'-t_b'}{2}\bigl(k_x^2-k_y^2\bigr)\,\tau_3\sigma_0.
\end{align}
To make the analysis simpler, we set $t=t_a'-t_b'$, which endows the Hamiltonian with an emergent $O(2)$ rotation symmetry. Although the results hold more generally, this simplification is enough to capture the emergence of nodes on the BZ boundary in the $d'$-wave state. Diagonalization of the Hamiltonian gives two parabolic bands: 

\begin{align}
E_1(\mathbf{k}) & = -\delta \mu - t'_a k^2 \\
E_2(\mathbf{k}) & = -\delta \mu - t'_b k^2
\end{align}
described by the eigenvectors 

\begin{align}
|u^{(1)}_{\uparrow,\downarrow}(\theta)\rangle & = (\cos\theta, \sin\theta)^T \otimes \ket{\up,\down}, \\
|u^{(2)}_{\uparrow,\downarrow}(\theta)\rangle & = (-\sin\theta, \cos\theta)^T \otimes \ket{\up,\down}
\end{align}
as a function of the angle $\theta$ around the $M$ point. Thus, the Bloch state describing the Fermi surface around the $M$ point is spin-degenerate and the corresponding pseudo-spinor undergoes a $4\pi$ winding going around the FS. For $\theta=0,\pi$, the FS state is completely  composed of electrons on the $A$ sublattice, and for $\theta=\pi/2,3\pi/2$, the FS state completely comes from the $B$ sublattice, as depicted in the inset of Fig. 1(c) of the main text.

As discussed in the main text, the $d'$-wave pairing vertex is given by $\Delta_d \sin \frac{k_x}{2} \sin \frac{k_y}{2} \tau_1\sigma_2$. Around the $M=(\pi,\pi)$ point, the explicit $\bm k$ dependence can be neglected, since $ \sin \frac{k_x}{2} \sin \frac{k_y}{2} \approx 1 $. However, when projecting the pairing gap onto the FS, an additional $\bm k$ dependence is generated, which follows from~\cite{WuPRB2023}
\be
\Delta_d^{(1,2)}(\bm k_\theta) = \langle u_{\uparrow}^{(1,2)}(\theta)| i\Delta_d\tau_1 \sigma_2 \left(|u_{\downarrow}^{(1,2)}(\theta+\pi) \rangle\right)^* = \pm \Delta_d \sin(2\theta). 
\ee

As can be seen from this simple calculation, even though the $d'$-wave form factor does not explicitly change sign around the $M$ point, the gap function projected onto the FS exhibits  $d_{xy}$-wave ($B_{2g}$) like nodes at $\theta = 0, \pm \pi/2, \pi$. It is straightforward to see that this results holds for generic values of $(t_a' - t_b')/t$.

While it is more common to associate $B_{2g}$ pairing with gap nodes along $k_x=0$ and $k_y=0$, we show that here the nodes located at the BZ boundaries are also enforced by symmetry. 
As we discussed in the main text, the BdG Hamiltonian $\mathcal{H}_{\rm BdG}(\bm k)$ and the Bloch wave function, upon shifting by a reciprocal lattice vector $\bm G_{1,2}$, satisfy
\be
\mathcal{H}_{\rm BdG}(\bk + \bm{G}_{1,2}) = \tau_3 \mathcal{H}_{\rm BdG}(\bk) \tau_3,~~~|u_{\uparrow,\downarrow}(\bm k + \bm{G}_{1,2}) \rangle = \tau_3  |u_{\uparrow,\downarrow}(\bm k ) \rangle.
\label{eq:nonsymm2}
\ee

It immediately follows that, when projected onto the FS, the pairing gap  $\Delta_d(
\bk_\theta)= \langle u_{\uparrow}(\bk_\theta)| i\Delta_d\tau_1 \sigma_2 \left(|u_{\downarrow}^{(1,2)}(\bk_{\theta+\pi}) \rangle\right)^*$, being a matrix element, is periodic upon a $\bm{G}_{1,2}$ shift, even though the \emph{pairing vertex} (e.g., $\Delta_d \sin \frac{k_x}{2} \sin \frac{k_y}{2} \tau_1\sigma_2$) may not be. As a result, the $M$ point is a fixed point for the action of $C_4$ on $\Delta(\bm k_\theta)$. Indeed, under $C_4$, $\Delta{((\pi, \pi))} \to \Delta{((-\pi, \pi))}\equiv \Delta{((\pi, \pi))}$, where in the last step we have used the periodicity of $\Delta(\bk_\theta)$ across the BZ. 
Therefore, as far as $\Delta_d(\bm k_\theta)$ is concerned, the $C_4$ rotation operation can be taken around the $M$ point (just as around $\Gamma$ point)
\be
C_4:~~~ \Delta_d(\theta) \to \Delta_d(\theta+\pi/2),
\ee
where we remind $\theta$ is the polar angle around $M$ point.

By the same token, one can show that under mirror reflections 
\begin{align}
\mathcal{M}_x:&~~~ \Delta_d(\theta) \to \Delta_d(\pi-\theta) \nonumber\\
\mathcal{M}_y:&~~~ \Delta_d(\theta) \to \Delta_d(-\theta).
\end{align}
Using the $B_{2g}$ ($d_{xy}$-wave) nature of $\Delta_{d}$, the pairing gap is odd under $C_4$, $\mathcal{M}_x$, and $\mathcal{M}_y$.
We thus conclude that on a single spin-degenerate FS, it has symmetry enforced nodes at $\theta = 0, \pm \pi/2, \pi$. 

The argument above applies both to our intra-unit-cell $d'$-wave order parameter, and to a more conventional $d$-wave pairing order with $\bar\Delta_{d}\sin k_x \sin k_y \tau_0 i\sigma_2$. In fact, even if $\Delta_d$ and $\bar\Delta_d$ are mixed, the nodes in the pairing gap remain robust.

\section{VII. Trilinear coupling between AM, $d'$-wave , and  $s'$-wave order parameters}

\begin{figure}
\includegraphics[width=0.4\columnwidth]{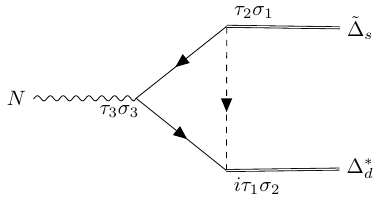}
\caption{The Feynman diagram coupling the AM order parameter $N$, the $d'$-wave order parameter $\Delta_d$ and a separate $s'$-wave order parameter $\tilde\Delta_s$. }
\label{fig:tri}
\end{figure}

Table \ref{table:SYM3} shows that symmetry allows for a trilinear term in the Landau free energy that couples the AM, SC $s$-wave, and SC $d$-wave order parameters:
\be
F_{3} = i \gamma N (\Delta_s \Delta_d^* - \Delta_d \Delta_s^*).
\label{eq:f30}
\ee
where $\Delta_s$ and $\Delta_d$ correspond to the $s'$- and $d'$-wave order  in the main text. However, a direct evaluation of the coefficient $\gamma$ shows that this term vanishes. Qualitatively, the reason is that both $\Delta_s$ and $\Delta_d$ are spin singlets, and the time-reversal symmetry breaking composite order, $i (\Delta_s \Delta_d^* - \Delta_d \Delta_s^*)$, corresponds to orbital currents. However, due to the lack of spin-orbit coupling, they cannot couple to the AM order parameter $N$, which is a pure spin order.

It turns out a similar trilinear coupling is nonzero for a different \emph{spin-triplet} $s'$-wave pairing order,  also discussed in Table \ref{table:SYM3}, which couples to the fermions as
\be
\tilde\Delta_s\, \sin\frac{k_x}{2} \sin\frac{k_y}{2} \psi_{\bm k}^\dag (\tau_2 \sigma_1) \psi_{\bm k}^{\dag T} .
\ee
As shown in Table \ref{table:SYM3}, $\tilde\Delta_s$ transforms as an $s'$-wave, the same as $\Delta_s$. Yet, due to its spin-triplet nature, the trilinear coupling 
\be
\tilde F_{3} = i \gamma N (\tilde\Delta_s \Delta_d^* - \Delta_d \tilde\Delta_s^*)
\label{eq:f3}
\ee
is nonzero.

To see this, we consider the Feynman diagram in Fig.~\ref{fig:tri}. The internal fermion lines are computed in the band basis, and the fermionic Green's functions are given by
\be
G (k) = G_c (k) + G_v (k) = -\frac{\ket{u(\bk)} \bra{u(\bk)}}{i\omega_m - \xi_c(\bm k)} -\frac{\ket{v(\bk)} \bra{v(\bk)}}{i\omega_m - \xi_v(\bm k)},
\ee
where 
\be 
\ket{u(\bk)} = (\cos \theta , \sin \theta)^T \otimes \ket{\up,\down}
\label{eq:63}
\ee
is the Bloch state of the conduction band in the vicinity of the $M$ point, and 
\be
\ket{v(\bk)} = ( \sin \theta,-\cos \theta)^T \otimes \ket{\up,\down}
\label{eq:64}
\ee
is that of the valence band.

In Fig.~\ref{fig:tri}, we focus on a specific arrangement in the band basis, where solid (dashed) lines correspond to the conduction (valence) band. Such an arrangement is necessary, since it can be directly verified that $\tilde\Delta_s$ projects to zero onto the conduction band (including onto the FS). In principle one also needs to consider the diagram with dashed and solid lines switched, but the result will be suppressed by the band gap.

The diagram in Fig.~\ref{fig:tri} is given by
\be
i\gamma = -T\sum_{\omega_m}\int_{\bk} \Tr\left[\frac{\ket{u(\bk)} \bra{u(\bk)}}{i\omega_m - \xi_c(\bm k)}\tau_3 \sigma_3 \frac{\ket{u(\bk)} \bra{u(\bk)}}{i\omega_m - \xi_c(\bm k)} \tau_2\sigma_1 \frac{\left(\,\ket{v(-\bk)} \bra{v(-\bk)}\,\right)^T}{-i\omega_m - \xi_v(-\bm k)} i\tau_1\sigma_2\right]
\ee
Evaluating the matrix elements around the $M$ point, we get
\be
\gamma = T\sum_{\omega_m}\int \frac{dk d\theta \cos^2 2\theta}{2\pi} \frac{ [i\omega_m +\xi_c(k)]^2(i\omega_m - \xi_v(k))}{(\omega_m^2+\xi^2_c(k))^2(\omega_m^2 + \xi^2_v(k))}.
\ee
The exact value of this integral depends on the fermionic dispersion, but it is straightforward to verify that $\gamma$ is real. As $\tilde F_3$ has to be real, it must have the form of Eq.~\eqref{eq:f3}.

Finally, as we mentioned in Sec. III.3, in the presence of SOC, $\Delta_s$ and $\tilde \Delta_s$ are always mixed (see the form factor $\Delta_6(k)$ in Fig.~\ref{fig:withSOC}). Via this mixing, the trilinear coupling in Eq.~\eqref{eq:f30} is also generated.

\bibliography{AlterM}